\documentclass{article}
\usepackage{epsfig}
\input epsf.sty
\newcommand{\D}{{\cal D}}
\newcommand{\la}[1]{\label{#1}}
 
\newcommand{\be}{\begin{equation}}
\newcommand{\ee}{\end{equation}}
\newcommand{\ba}{\begin{eqnarray}}
\newcommand{\ea}{\end{eqnarray}}
\newcommand{\bi}{\begin{itemize}}
\newcommand{\ei}{\end{itemize}}
\newcommand{\rmi}[1]{{\mbox{\scriptsize #1}}}
\newcommand{\nr}[1]{(\ref{#1})}

\newcommand{\nn}{\nonumber \\}
\newcommand{\fr}[2]{{\frac{#1}{#2}}}
\newcommand{\msbar}{\overline{\mbox{\rm MS}}}

\newcommand{\bfx}{{\bf x}}

\newcommand{\eq}{Eq.~}
\newcommand{\eqs}{Eqs.~}
\newcommand{\fig}{Fig.~}
\newcommand{\figs}{Figs.~}
\newcommand{\se}{Sec.~}
\newcommand{\half}{{1\over2}}

\def\lsi{\raise0.3ex\hbox{$<$\kern-0.75em\raise-1.1ex\hbox{$\sim$}}}
\def\gsi{\raise0.3ex\hbox{$>$\kern-0.75em\raise-1.1ex\hbox{$\sim$}}}

\newcommand{\gsim}{\mathop{\gsi}}
\makeatletter \@addtoreset{equation}{section} \makeatother

\begin{document}
 
\begin{titlepage}
\begin{flushright}
CERN-TH/99-168\\
NORDITA-99/24HE\\
HIP-98-085/TH\\
hep-lat/9906028\\
\end{flushright}
\begin{centering}
\vfill
 
{\bf STATISTICAL MECHANICS OF VORTICES \\
     FROM FIELD THEORY}

\vspace{0.8cm}

K.~Kajantie$^{\rm a,}$\footnote{keijo.kajantie@helsinki.fi},
M.~Laine$^{\rm b,a,}$\footnote{mikko.laine@cern.ch},
T.~Neuhaus$^{\rm c,d,e,}$\footnote{neuhaus@physik.rwth-aachen.de},
A.~Rajantie$^{\rm f,}$\footnote{a.k.rajantie@sussex.ac.uk} and
K.~Rummukainen$^{\rm g,d,}$\footnote{kari@nordita.dk} \\

\vspace{0.3cm}
{\em $^{\rm a}$Dept.\ of Physics,
P.O.Box 9, FIN-00014 Univ.\ of Helsinki, Finland\\}
\vspace{0.3cm}
{\em $^{\rm b}$Theory Division, CERN, CH-1211 Geneva 23,
Switzerland\\}
\vspace{0.3cm}
{\em $^{\rm c}$Institut f\"ur Theoretische Physik E, RWTH Aachen, FRG\\}
\vspace{0.3cm}
{\em $^{\rm d}$Helsinki Inst.\ of Physics,
P.O.Box 9, FIN-00014 Univ.\ of Helsinki, Finland\\}
\vspace{0.3cm}
{\em $^{\rm e}$ZiF, Univ.\ Bielefeld, D-33615 Bielefeld, FRG\\}
\vspace{0.3cm}
{\em $^{\rm f}$Centre for Theor. Physics, Univ. of Sussex, 
Brighton BN1 9QH, UK\\}
\vspace{0.3cm}
{\em $^{\rm g}$NORDITA, Blegdamsvej 17,
DK-2100 Copenhagen \O, Denmark\\}

\vspace{0.7cm}
{\bf Abstract}
 
\end{centering}
 
\vspace{0.3cm}\noindent
We study with lattice Monte Carlo simulations
the interactions and macroscopic behaviour of a large 
number of vortices in the 3-dimensional U(1) gauge+Higgs field 
theory, in an external magnetic field. 
We determine non-perturbatively the (attractive or repelling) interaction 
energy between two or more vortices, as well as the critical 
field strength $H_c$, the thermodynamical discontinuities, 
and the surface tension related to the boundary between the 
Meissner phase and the Coulomb phase in the type~I region.
We also investigate the emergence of vortex lattice and 
vortex liquid phases in the type~II region. For the type I region
the results obtained are in qualitative agreement with mean field 
theory, except for small values of $H_c$, while in the type II region
there are significant discrepancies. These findings are relevant 
for superconductors and some models of cosmic strings, as well as 
for the electroweak phase transition in a magnetic field.
\vfill
\noindent
 
\vspace*{1cm}
 
\noindent
CERN-TH/99-168\\
NORDITA-99/24HE\\
HIP-98-085/TH\\
September 1999
 
\vfill

\end{titlepage}
 
\section{Introduction}

The continuum 3-dimensional (3d) U(1) gauge+Higgs 
field theory (scalar electrodynamics) 
is of interest for several reasons. First of all, it is
the Ginzburg-Landau (GL) theory of superconductivity, 
believed to be applicable for low-$T_c$ and possibly
also for some aspects of high-$T_c$ superconductors. Second, 
it is an often-used toy model for studying line-like 
cosmological topological defects, strings
(adopting here the condensed matter terminology, we 
will mostly use the word vortex for these defects). 
Finally, the existence of topological excitations makes the model
interesting also as a theoretical playground: much of the
terminology related to our understanding of confinement, 
for instance, arises from superconductors.
It is rather remarkable that for this theory, the topological observables 
of the continuum limit can be easily generalized in a gauge invariant
way to the case with a finite lattice cutoff~\cite{ref:old,givortex}.

In all these different contexts, it is of interest to ask
how the behaviour of the system changes when an external 
magnetic field is added. In superconductors, an external 
magnetic field can be imposed experimentally, and it leads to 
the emergence of completely new phases, displaying 
for instance broken translational invariance. Due to magnetohydrodynamic
diffusion, essentially homogeneous magnetic fields could also exist
in cosmology and be relevant for the phase transitions appearing there; 
in any case, the core of each long U(1) vortex carries magnetic flux. 
{}From the theoretical point of view, a small external
magnetic flux allows also to probe the topological properties 
of the zero-flux system in a gauge-invariant and systematic way. 

In a previous paper~\cite{tension}, we showed how one unit
of external magnetic flux can be imposed gauge-invariantly
on the lattice, and how it can be used to determine
non-perturbatively the free energy per unit length, i.e. tension, 
of a long vortex. 
The vortex tension thus defined was shown
to constitute a gauge-invariant (but non-local) 
order parameter for the system, 
differentiating between the symmetric (``Coulomb'', ``normal'')
and broken (``Meissner'', ``Higgs'', ``superconducting'') phases. 
In the present 
paper, we study the case of a larger magnetic field. We show
how to measure the free energy of a system of two or more vortices, 
thus seeing whether vortices attract or repel each other. This
allows to divide the phase diagram into the region of type I 
and type II superconductors. We also increase
the magnetic field further on and inspect whether this 
can lead to the emergence of new phases, triggered by the 
interactions of vortices, 
which are themselves macroscopic objects
from the point of view of the original continuum 
quantum field theory. 

In condensed matter physics vortices, of course, are a thoroughly
studied topic (see, e.g., \cite{blatter}--\cite{nat}). 
The emphasis is somewhat different, 
though. There the existence of a multitude of different scales and
parameters implies that even the starting point, theory or model,
is not uniquely known. The multitude of scales is related to the fact
that one usually stays away from the immediate vicinity of the 
transition point, as a result of which the mean field approximation 
also often works very well. In particular, 
if one is studying the GL theory, one often approximates it
by a simpler theory (the frozen gauge model, the XY model, etc.), 
using the narrowness of the fluctuation region as an argument. These
arguments may be correct for physics, as it is only experiment
which decides which effective theory is best, but they are not correct 
for the GL theory as such.

In the particle physics context, in contrast, given definite
symmetry principles and the requirement of ultraviolet insensitivity, 
the form of the 
theory is uniquely defined; in the case of the U(1)+Higgs theory, 
keeping only relevant operators,  
it depends just on two dimensionless parameters.
Moreover, close enough to the transition point, 
mean field approximation completely fails, and the theory
is a genuine quantum field theory. It is quite striking that even for 
a simple theory such as U(1)+Higgs, 
the properties of the phase diagram are not yet
completely known. We hope that studying the response of the system
to an external magnetic field will shed new light on this issue.

Somewhat similar topics (with different methods and in 4d)
were first studied with lattice simulations in~\cite{dh}. 
External magnetic fields in the 3d SU(2)$\times$U(1) theory 
(corresponding to the
Standard Model) were studied in~\cite{bext}, and in
the 3d U(1) theory with fermions 
instead of scalars, in~\cite{farakos}.

The plan of the paper is the following. In \se\ref{theory}
we define the theory in the continuum and on the lattice, and
review how an external magnetic field can be imposed and how
the free energy related to vortices can be measured.  
In \se\ref{meanfield} we review the mean field results for the
vortex and surface tensions and for the behavior of the system in
an external magnetic field.
In \se\ref{phasediag} we consider the full path integral and
discuss the validity of the mean field approximation.
In \se\ref{sec:typeI} we present our simulation results for
the type I region, and in \se\ref{sec:typeII} for the type II region.  
We conclude in \se\ref{concl}.

\section{An external magnetic field and  vortices
on the lattice}
\la{theory}


\paragraph{The theory in continuum without external field.}
The theory underlying our considerations is the
U(1)+Higgs theory, corresponding to high temperature
scalar electrodynamics on one hand~\cite{perturbative}--\cite{joa} 
and to the  
Ginzburg-Landau model of superconductivity on the other~\cite{kleinert}.
It is defined by the functional integral
\ba
Z&=&\int {\cal D}A_i{\cal D}\phi\,\exp\bigl[-S(A_i,\phi)\bigr], \la{z} \\
S&=&\int d^3x\biggl[\fr14 F_{ij}^2+
|D_i\phi|^2 
+m_3^2 \phi^*\phi + \lambda_3 \left(\phi^*\phi\right)^2\bigg], 
\label{action}
\end{eqnarray}
where $F_{ij}=\partial_iA_j-\partial_jA_i$ 
and $D_i=\partial_i+ie_3A_i$. When all dimensionful quantities
are expressed in proper powers of the scale $e_3^2$, 
the theory can be parametrized by the two dimensionless ratios
\be
y= { m_3^2(e_3^2)\over e_3^4},\quad x={\lambda_3\over e_3^2},
\la{parameters}
\ee
where $m_3^2(\mu)$ is the mass parameter in 
the $\msbar$ dimensional regularization scheme in 
$3-2\epsilon$ dimensions. For completeness, one may note
that the standard dimensionful textbook coefficients $a,b$~\cite{lp}
of the Ginzburg-Landau free energy and the dimensionless GL parameter
$\kappa$ are related to $y,x$ by 
\be
y= {mc^2\over64\pi^2\alpha^2T^2}a,\quad 
x={1\over2\pi\alpha\hbar c}\left({mc\over\hbar}\right)^2b=
\kappa^2,
\la{ab}
\ee
where $\alpha$ is the fine structure constant and 
$m$ is an effective mass parameter. Note the huge
numerical ratios entering here.

\paragraph{The theory on the lattice without external field.}
The discretized 
lattice action corresponding to \eq\nr{action} is 
\ba
S[\alpha,\phi] & = & \beta_G \!\!\sum_{\bfx,i<j}
\fr12\alpha_{ij}^2(\bfx)
  -{2\over\beta_G} \sum_{\bfx, i} {\mbox{Re}}\, 
\phi^*(\bfx) U_i(\bfx)\phi(\bfx+\hat{i})\nonumber\\   
  & & +\beta_2\sum_{\bfx} \phi^*(\bfx) \phi(\bfx)
+{x\over\beta_G^3}\sum_{\bfx} \left[\phi^*(\bfx)\phi(\bfx)\right]^2, 
 \la{standardaction} 
\ea
where $\alpha_i(\bfx) = a e_3 A_i(\bfx)$,
$\alpha_{ij}(\bfx)=\alpha_i(\bfx)+\alpha_j(\bfx+\hat i)-
\alpha_i(\bfx+\hat j)-\alpha_j(\bfx)$, $U_i(\bfx)=\exp[i\alpha_i(\bfx)]$,
and where the lattice couplings $\beta_G, \beta_2$ are 
related to the continuum parameters and the lattice constant
$a$ by~\cite{contlatt} (for ${\cal O}(a)$-corrections, see~\cite{moore_a})
\ba
& & \beta_G={1\over e_3^2a},
\la{betaR} \\
& & \beta_2={1\over\beta_G}
\biggl[6+{y\over\beta_G^2}
-{3.1759115(1+2x)\over2\pi\beta_G}\la{betah}\nonumber \\
&&\hspace*{1cm} -\frac{
(-4+8x-8x^2)(\ln 6\beta_G+0.09)-1.1+4.6x}{16\pi^2\beta_G^2}\biggr].
\ea
Note that we must use here the non-compact formulation for the
U(1) gauge field, to avoid topological artifacts (monopoles) at 
finite $a$ (with the compact action one needs
to go closer to the continuum limit to start with
than $\beta_G\ge 1$ as we have here, 
at least to $\beta_G\ge 4$; see~\cite{u1big}).

\paragraph{The thermodynamical ensembles with an external field.}
What we do in this paper is an extension to $n$ vortices
of what was done for one vortex in \cite{tension}. Physically,
we impose an external magnetic flux $\Phi_B$
through the volume, 
\be
e_3 \Phi_B \equiv e_3 BL_1L_2 = 2\pi n. 
\la{flux}
\ee
By convention, we have fixed the flux density as ${\bf B}=(0,0,B)$, 
and use an $N_1N_2N_3$ lattice with $L_i=N_ia$, $V=L_1L_2L_3$, $A=L_1L_2$.
On the average, after taking into account translational invariance, 
this flux density determines that 
$\langle F_{12} \rangle = B$, although in individual configurations
the flux may not be evenly distributed. 
Together with the requirement of periodicity of observable quantities, 
\eq\nr{flux} guarantees that there are $n$ 
vortices passing through the system, and we 
want to measure the free energy $F(B)$ of such a configuration.
Technically, fixing the flux is equivalent to multiplying
the integrand in \eq\nr{z} by the delta-function
\be
\Pi_{x_3=1}^{N_3}
\delta\Bigl(\int dx_1dx_2 e_3 F_{12}(\bfx) - 2\pi n \Bigr),
\ee
where the product has been written in a discretized form.

We shall use interchangeably the notation $F(n)$ or $F(B)$ for
the free energy thus obtained, noting that $n$ and $B$ are related by
\eq\nr{flux} and that $F$ also depends on the parameters $e_3^2,y,x$
(see \eq\nr{parameters}) of the theory. Note that 
in our convention the free energy is dimensionless, 
$Z=\exp(-F)$, so that it
corresponds to the free energy divided by temperature in standard
terminology. 

In a thermodynamic sense, fixing the flux means 
choosing a microcanonical ensemble.
The results can also be transformed
to the canonical ensemble $G(H)$,
\be
G(H)=F(B)-VHB,\quad F'(B)=VH, \la{canonical}
\ee
where $V$ is the volume of the system and $H$ is the external 
field strength. Technically, the canonical ensemble means
multiplying the integrand in \eq\nr{z} by the factor
\be
\exp\bigl[ + H \int d^3 x F_{12}(\bfx)\bigr]. \la{canon}
\ee

\paragraph{Imposing the magnetic flux on the lattice.}
To impose a flux $\Phi_B$
of $n$ units of $2\pi/e_3$ on the lattice, we choose here to 
proceed as follows. We 
define a modified theory by
\be
\label{equ:freeen}
e^{-F(m)}=\int\D\alpha\D\phi \,\,
e^{-S[\alpha,\phi;m]},
\ee
where the degrees of freedom are the periodic link angles
\be
-\infty<\alpha_i(\bfx)<\infty,\quad \alpha_i(N_1+1,x_2,x_3)=
\alpha_i(1,x_2,x_3), \,\,{\rm etc.},
\la{periodic}
\ee
the periodic scalars $\phi(\bfx)$, and
\ba
\label{equ:actchange}
S[\alpha,\phi;m]&=&\beta_G\sum_{\bfx}
\fr12[(\alpha_{12}(\bfx)+2\pi m\delta_{x_1,x_0}\delta_{x_2,y_0})^2+
\alpha_{23}^2+\alpha_{13}^2]+... \nonumber \\
&=&S[\alpha,\phi]
+\beta_G\sum_{x_3}(2\pi m\alpha_{12}(x_0,y_0,x_3)+2\pi^2 m^2).
\ea
Since, to leading order in $a\to0$, 
\be
\alpha_{12}(x_0,y_0,x_3)=a^2 e_3 F_{12}(x_0,y_0,x_3)=e_3\Phi_B,
\ee
we have forced a flux of $-2\pi m/e_3$ through the lattice; the stack
of plaquettes parallel to the $x_3$ axis at the position $(x_0,y_0)$
is, in fact, a ``Dirac string'' carrying the flux $2\pi m/e_3$ in
the $-x_3$ direction.

A crucial fact now is that, due to the periodicity \nr{periodic},
\be
\sum_{x_1,x_2}\alpha_{12}(x_1,x_2,x_3)=0;\quad{\rm any}\,\,x_3,
\la{fluxconservation}
\ee
and thus the total flux through any plane vanishes identically.
This implies that the flux $+2\pi m/e_3$ must return through
the system in the $+x_3$ direction but now in a manner specified
by the dynamics of the theory. This response is the object of 
the study here.

The Dirac string has been imposed by modifying one special plaquette
on each $x_3$ plane leaving the part of the action involving scalars
unchanged. When $m$ is an integer, integrating over all periodic
field configurations makes the path integral (\ref{equ:freeen})
translation invariant.
One could also, equivalently, impose the flux by making a non-periodic 
change in the boundary conditions for the 
link variable $\alpha_i$, as was done in Ref.~\cite{tension}.
Although translation invariance is broken with non-integer $m$,
the path integral (\ref{equ:freeen}) is well-defined,
and non-integer values will indeed be used to interpolate between
integers.

Finally, in principle one might also attempt to perform simulations
with the canonical ensemble, as was done in~\cite{dh}.  In our case
this can be accomplished by coupling an external field to the
flux according to \eq\nr{canon}, i.e., adding a term 
$H L_3 \int d^2 x F_{12}(\bfx) = H L_3 2\pi n /e_3$
to the action. This would promote $n$ to a dynamical 
variable, for which only integer values are allowed.   
However, in our case it would be very difficult to obtain 
an efficient update for $n$. This is because the update is 
of semi-global nature: by performing it one attempts to make 
a large change in a configuration without any interpolation.

\paragraph{Measuring the free energy.}
The aim now is to measure the change in the free energy caused
by switching on the magnetic field, $F(n)-F(n=0)$.
Since one cannot measure the absolute value of 
the free energy on the lattice,
we eliminate the unknown $F(0)$ 
by taking a derivative with respect to $m$ and integrating back.
This leads to the following result for the free energy per 
unit length (always relative to $F(0)$):
\ba
\label{F(N)}
{F(n)\over L_3e_3^2}&=&2\pi^2 \beta_G^2 \int_0^n dm \,W(m) \nn
&\equiv& 2\pi^2 \beta_G^2 \int_0^n dm
\left[2m+{1 \over {\pi N_3}} \Bigl\langle \sum_{x_3} \alpha_{12}(x_0,y_0,x_3)
\Bigr\rangle_m \right].
\ea
The subscript $m$
emphasises the fact that the expectation value is computed using the
action \nr{equ:actchange} with the defect.
We recall that we have chosen to express every dimensionful 
quantity in units of an appropriate power of $e_3$. For instance,
$F(n)/L_3$ was made dimensionless in \eq\nr{F(N)}
by dividing it by $e_3^2$.

Since $W(m)$ is the quantity measured in our simulations, it is useful
to have a feeling of its magnitude. It is essentially the expectation
value of the stack of plaquettes at the position of the Dirac string.
Eq.~\nr{equ:actchange} for the first implies that the action is small
for $\alpha_{12}(x_0,y_0,x_3)=-2\pi m$. This large negative
contribution cancels the term $2m$ in \nr{F(N)}. This rough estimate
can be improved by evaluating $W(m)$ including only the gauge field
in the action and using \nr{fluxconservation}. The plaquettes 
$\alpha_{12}$ are of two types: one at the string and $N_1N_2-1$ not
at the string. Using shift symmetries of the action one can then
prove that
\be
W(m)={2 m\over N_1N_2},
\la{wmappro}
\ee
for all $m$, which simply corresponds to $F=\fr12 VB^2$, $B$ given by
\nr{flux}. In fact, the shift symmetries
also hold for the full theory if $m=n$ = integer, so that 
\eq\nr{wmappro} will give $W(n)$ exactly when
$n$ = integer. The behaviour of $W(m)$ between integer values 
is seen from the numerical computations in 
\figs\ref{WtypeI}(a-c).

\paragraph{Measuring the field strength.}
In addition to the total free energy of $n$ flux units, 
we will be interested in the increase of free
energy when the flux is increased by one unit, $dF(n)\equiv
F(n)-F(n-1)$. Using $dB\cdot L_1L_2=2\pi /e_3$ and \eq\nr{canonical},
this is, in fact, 
a measurement of $H$:
\be
\frac{H}{e_3^3}=
\frac{1}{e_3^3V}\frac{dF}{dB}=\frac{1}{2\pi}\frac{dF}{L_3e_3^2}
=\pi \beta_G^2 \int_{n-1}^n dm W(m).
\la{dF}
\ee


\begin{figure}[t]

\centerline{\psfig{file=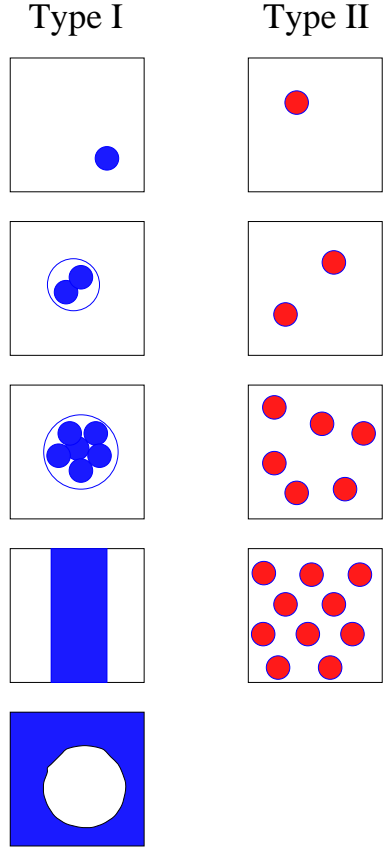,angle=270,width=5cm}}
\caption{The behaviour of the intersection of 
flux lines with increasing total flux 
on a finite lattice in type I and II cases (see text).
A dark area is a region of the symmetric phase
carrying the flux, while a white area is
a region of the broken phase without a flux. 
The notation is somewhat symbolic: in the type I case,  
the flux lines are in reality 
on top of each other.}
\la{fluxincrease}
\end{figure}

\section{Mean field results and \\
the determination of physical observables}
\la{meanfield}

In this section, we discuss the behaviour 
of the system at mean field level, and how this pattern
can be employed in determining some of the physical observables 
of the full quantum theory. 

\paragraph{One or two vortices.}
In the mean field approximation the phase of the system is determined by the
configuration that minimizes the action \nr{action} under
the boundary conditions imposed. In the bulk, 
for zero flux the system has two phases depending on $y$; a broken phase
at $y<0$ and a symmetric phase at $y>0$. The transition between the
phases is of second order. In the broken phase the
scalar and  vector masses are
\be
m_S^\rmi{MF}=\sqrt{-2y}e_3^2,\quad m_V^\rmi{MF}=\sqrt{-y\over x}e_3^2.
\la{mfmasses}
\ee

Assume now that the whole
system is in the broken phase and ask what happens if one starts
increasing the magnetic flux in the $x_3$ direction keeping the
transverse area fixed (\fig\ref{fluxincrease}). This is
concretely how our simulations will be organised.

Taking first
$n=1$ one vortex line appears somewhere. 
By definition, 
the vortex tension $T$ is the free energy of this configuration
divided by the length of the system in the $x_3$-direction.
In the mean field approximation,
$T$ can be calculated by minimizing the action \nr{action} with
cylindrically symmetric boundary conditions. The result is
\cite{jr}
\be
T={F(n=1)\over L_3} = e_3^2\left({-y\over x}\right)\pi{\cal E}(\sqrt{2x})\equiv
T^\rmi{MF},
\la{T}
\ee
where ${\cal E}(1)=1$, ${\cal E}(0.3043)=0.63$ ($x=0.0463$) and
${\cal E}(2)=1.32$ ($x=2$).

When $n$ is increased to
$n=2$, a difference
between type I ($x<1/2$) and type II ($x>1/2$) appears: in the
former case vortex lines attract and in the latter they repel
each other. If we form the quantity
\be
T_2={F(n=2)\over 2L_3},
\la{T2}
\ee
we thus expect that for type I,
\be
\la{type1}
T-T_2>0 \quad ({\rm type\, I}),
\ee
due to the binding energy of two vortices. This also holds in the
thermodynamic limit. For type II, 
\be
\la{type2}
T-T_2\le0 \quad ({\rm type\, II}),
\ee
where the inequality holds on finite lattices and the 
equality in the thermodynamic limit. Thus $T-T_2$ is an order
parameter for separating the type I and type II regions. 
In \cite{tension} it has been shown that $T$ itself, on the other hand,  
serves as an order parameter for the normal $\leftrightarrow$ 
superconducting phase transition in this theory: 
it vanishes in the thermodynamic limit for $y>y_c(x)$. 

The boundary between type I and II, $x=1/2$, is mathematically very
interesting since the classical equations of motion then simplify 
considerably. The dynamics of $n$ vortices 
can then be largely solved~\cite{manton}. We shall not study this
special case here.

\subsection{Type I, $x<1/2$}
\la{typeImf}

\paragraph{The general pattern.} 
Consider then type I and increase $n$ further. The lines 
parallel to $x_3$ first cluster on the $x_1,x_2$ plane as a 2d 
bubble (= 3d cylinder) of symmetric phase, but at some 
stage it becomes more favourable to form a
slab (see \fig\ref{fluxincrease}). The cylinder and slab have
interfaces separating the broken and symmetric phases. 
The tension of this interface can be written as
\be
\sigma\equiv\sigma^\rmi{MF}={(-y/2)^{3/2}\over x} s(x) e_3^4,
\la{sigmainterface}
\ee
where $s(x)$ is the extremal value of an integral already written
down by Ginzburg and Landau. For small $x$ \cite{sigma},
\be
s(x) = \fr43-1.24x^{1/4}+...
\ee
and one can analytically prove that $s(x)$ vanishes for $x=1/2$ and
becomes negative for $x>1/2$. For intermediate values numerical
methods have to be used \cite{sigma,th}. For the value
of $x$ used in the type I simulations we find $s(x=0.0463)=0.643$.

With continuously increasing $n$ the slab gets thicker. The broken phase
region gets smaller and, ultimately, it reduces to 
a 3d cylinder of the broken phase, a hole in the magnetic field
configuration. At some critical value of the flux, 
corresponding to a critical value of the field strength $H_c$, 
the system goes entirely to the symmetric phase.

\paragraph{The detailed finite volume behaviour.}
Consider now an inhomogeneous configuration in which the area
$A_s$ is in the symmetric phase, and the area
$A-A_s$ is in the broken phase with no $B$. 
Let us denote the broken phase action density as
\be
{-m_3^4\over 4\lambda_3}\equiv-\fr12 H_c^2,\quad H_c\equiv
H_c^\rmi{MF}={-y\over\sqrt{2x}}e_3^3.
\la{hcmf}
\ee
In the symmetric phase, the 
magnetic flux density is $B_s=BA/A_s$,
and the action density is $\fr12 B_s^2$. If
$\partial A_s$ is the length of the boundary, the 
action per length is
\be
\frac{F}{L_3}
=\fr12A_sB_s^2+(A-A_s)(-\fr12H_c^2)+\sigma\partial A_s,
\la{dropaction}
\ee
where the three stages discussed above correspond to
\be\begin{array}{lrcl}
\mbox{cylinder of symmetric phase:} &
\quad A_s&=&\pi r^2,\,\,\partial A_s=2\pi r,\nn
\mbox{slab:} & \quad A_s&=&L_2d,\,\,\partial A_s=2L_2,\nn
\mbox{cylinder of broken phase:} &
\quad A_s&=&A-\pi r^2,\,\,\partial A_s=2\pi r.
\end{array}\ee
These now have to be minimised for cylinder radii $r$ and the slab
width $d$ (it was assumed that $L_2\le L_1$)
under the condition $\Phi=BA=B_sA_s$. The configuration corresponding
to the minimum of $F/L_3$ then is the stable one. The minimisation is
simplest for the slab: then
\be
d={B\over H_c}L_1.
\ee

For the cylinders one has to minimise numerically, in general. 
However, an analytic result is obtained if
the surface term is so small that it can be
neglected in the minimisation, though it affects the value of 
$F/L_3$. This is true if $\sigma\ll H_c^2r$ at the minimum value of $r$.
Then
\be
{F_\rmi{cyl.symm.}\over L_3}=\left(H_cB-\fr12 H_c^2\right)L_1L_2+2\sigma
\sqrt{\frac{\pi BL_1L_2}{ H_c}}.
\la{Sbubble}
\ee
For a slab configuration with $L_2\le L_1$, one obtains
\be
{F_\rmi{slab}\over L_3}=\left(H_cB-\fr12 H_c^2\right)L_1L_2+2\sigma L_2.
\la{Sslab}
\ee
Finally, for large $n$ one has a cylinder of broken phase within symmetric
phase and
\be
{F_\rmi{cyl.brok.}\over L_3}=\left(H_cB-\fr12 H_c^2\right)L_1L_2+2\sigma
\sqrt{\pi L_1L_2\Bigl(1-{B\over H_c}\Bigr)}.
\la{Sbubble3}
\ee
Comparing \nr{Sbubble} and \nr{Sslab} one sees that a cylinder is
favoured if
\be
\la{bubbleslab}
{B\over H_c}<{L_2\over \pi L_1}\equiv \frac{B_1}{H_c}.
\ee
To see when the transition to the homogeneous symmetric phase
takes place, one should compare with its free energy:
\be
{F_s\over L_3}=\fr12 AB^2.
\la{Ssymm}
\ee
One sees, comparing \nr{Sslab} and \nr{Ssymm},
that this is smaller than even the slab free energy if
\be
B>H_c-2\sqrt{\sigma\over L_1}\equiv B_2.
\la{critB}
\ee
The transition to the symmetric phase will take place directly
from the slab stage if the free energy of the 2nd cylinder stage is
at $B_2$ larger than that of the slab:
\be
{H_c\over2\pi}<{L_1\over L_2}\sqrt{\sigma\over L_1}
\la{skipcondition}
\ee
Our simulations will, in fact, satisfy this condition. To see the
stage with the cylindrical hole in the magnetic field configuration,
one has to go to still larger lattices. This result is also confirmed
by a full numerical minimization of \eq\nr{dropaction}.

\paragraph{The determination of $H_c$ and $\sigma$.}
On the basis of the above equations we thus have a definite scenario
for what happens when we increase the flux $n$ further from $n=2$
at finite volumes, and how this is related to the physical 
properties of the transition in the thermodynamical limit.

To see what \eq\nr{Sbubble} predicts for the measured
$H(B)$ (\eq\nr{dF}), simply take a derivative of  \eq\nr{Sbubble}
with respect to $B$ (we recall that the relation 
of $B$ and $n$ is as in \eq\nr{flux}).
Then, for small $B$ the measurements should behave as
\be
\la{dFbubble}
H(B)=
H_c+\sigma\sqrt{\frac{\pi}{L_1L_2BH_c}}.
\ee
For $B>H_cL_2/\pi L_1=B_1$ (cf.\ \eq\nr{bubbleslab}),
the slab configuration dominates and leads to a constant plateau, 
\be
\la{dFslab}
H(B)=H_c.
\ee
Finally, when $B=B_2$, the system enters the Coulomb phase in which
\be
\la{dFcoul}
H(B)=B.
\ee
Thus, we can determine $H_c$ from the plateau
according to \eq\nr{dFslab}, and then $\sigma$
from \eq\nr{dFbubble}.

\paragraph{The determination of thermodynamical discontinuities.}
We have here considered a system with a fixed $B$, on which
the magnetic field $H$ depends, but it may be more natural to move
to the canonical ensemble in which $H$ is fixed and $B$ is allowed
to fluctuate. This means simply
inverting the relation of $B$ and $H$. From this point of view, we simply
have a standard first-order phase transition at $H=B_2$, and when
$H$ is below that, $B$ vanishes. The configurations with a cylinder
or a slab describe the system on the transition line.

The thermodynamical discontinuities related to the first-order
transition are directly given by the properties of 
the canonical free energy $G(x,y,H/e_3^3)$. 
The function $G(x,y,H/e_3^3)$ is continuous across 
the phase transition, but its derivatives are not:
\be
\Delta \frac{\partial G/V}{\partial y} = e_3^4 \Delta
\langle \phi^*\phi\rangle, \quad
\Delta \frac{\partial G/V}{\partial H/e_3^3} = - e_3^3 \Delta B.
\la{disc}
\ee
For fixed $x$, the latent heat $L$ of the transition is 
defined as the 
discontinuity in the ``energy'' variable~$E$, obtained from 
$G$ by a Legendre transformation with respect to $y, H/e_3^3$:
\be
L = \Delta E  =  
-y \Delta \frac{\partial G}{\partial y} 
-H_c \Delta \frac{\partial G}{\partial H} =  
V \Bigl[
-y e_3^4  \Delta \langle \phi^*\phi\rangle + H_c \Delta B \Bigr].  
\la{latheat}
\ee
Note that for fixed $x$, the identity $\Delta G(x,y,H_c)=0$
leads to the Clausius-Clapeyron equation, relating
the different discontinuities:
\be
\Delta \frac{\partial G}{\partial y}= -
\frac{d H_c}{dy} \Delta \frac{\partial  G}{\partial H} 
\quad \Leftrightarrow \quad
e_3^4 \Delta \langle \phi^*\phi\rangle = 
\frac{d H_c}{dy} \Delta B.
\la{CC}
\ee
Thus it is enough to measure one of the discontinuities, 
and the curve $H_c(y)$. 
Below we choose to discuss $\Delta B$ (on finite volumes, 
$\Delta B = B_2$ instead of $\Delta B = H_c$).
Finally, let us recall that at the mean field level 
\eq\nr{CC} is satisfied
through 
\be
\Delta \langle \phi^*\phi\rangle = \Bigl(\frac{y}{2x}\Bigr)e_3^2 <0,\quad
H_c = -\frac{y}{\sqrt{2x}} e_3^3, \quad
\Delta B = H_c. 
\ee

\subsection{Type II, $x>1/2$}
\la{sub:typeII}

\paragraph{The determination of vortex tension and vortex interaction
energy.} In the type I region, the vortex tension can be determined 
directly from \eqs\nr{T}, \nr{T2}. In the type II region, the
issue is more involved, since there are large finite volume 
effects. Indeed, as discussed after \eq\nr{type2}, the 
difference between $T,T_2$ vanishes in the thermodynamical 
limit. However, this finite volume behaviour is well understood, 
and can be employed for the measurement of the
properties of the vortex system.

The basic idea is that due to the repelling interaction, the
vortices tend to form a vortex lattice in the type II region. 
In fact, due to the 
periodic lattice boundary conditions, even one vortex
forms a square lattice with its periodic counterparts. Thus
one can use finite size scaling with $n=1$ to study the interaction
energy of two vortices. The method can be continued with $n=2$, etc.
Of course, this is not equivalent to a real vortex lattice, in which
the distances between the vortices fluctuate; here the distance is
fixed by the lattice size. When $n$ grows to much larger values
so that the vortex separation is no longer determined by the lattice 
size, the vortices are instead supposed to form
a physical (triangular) Abrikosov lattice.

Mean field theory offers an ansatz for the interaction energy:
if $d\gg 1/m_V$ 
is the physical distance between two vortices (it can be the
lattice size $L$ in finite size scaling studies of $T$ or $T_2$ or
the average distance between vortices,
$d\approx \sqrt{L_1L_2/n}=\sqrt{2\pi/(Be_3)}$, at large $n$),
then the interaction energy is~\cite{br}
\be
\epsilon_{12}=2 T X K_0(m_Vd),
\la{vortexinteraction}
\ee
where $X=1/{\cal E}(\sqrt{2x})$,  
$K_0(z)$  is a Bessel function ($\sim \sqrt{\pi/(2z)}\exp(-z)$ for 
$z\to\infty$), and $m_V$ is the photon mass (in the
broken phase). We have chosen to parametrize the prefactor in terms 
of the tension $T$ for later notational convenience, although, 
in fact, the physics of the prefactor is different from
that in $T$: the prefactor is only sensitive to asymptotic fields
and the photon mass, and thus ${\cal E}(\sqrt{2x})$
cancels between the mean field expressions for $T$ and $X$.
Let us also recall that the function $K_0$ appears 
in \eq\nr{vortexinteraction} because it determines 
the profile of a single vortex. 

In the following, we will need an ansatz for the
total free energy of a system of vortices forming either 
a square or a triangular lattice. We define the lattice sums
\ba
S_S(z)&=&\sum_{n_i=-\infty}^\infty K_0\left(z\sqrt{n_1^2+n_2^2}\right),\nn
S_T(z)&=&\sum_{n_i=-\infty}^\infty K_0\left(z\sqrt{(2/
\sqrt3)(n_1^2+n_2^2+n_1n_2)}\right),
\la{lattsums}
\ea
where the term $n_1=n_2=0$ is excluded and where the factor $2/\sqrt3$ 
follows by requiring that the two lattices have the same number of
sites within a large rectangular area. (From the point of view
of a triangular lattice, it would be more natural to use
a lattice geometry which is not rectangular but has $L_1:L_2\approx
2:\sqrt{3}$, but we stick here to $L_1=L_2$.)
A numerical computation of
the sums shows that they are very close to each other in the
relevant range of $z$; this is the known fact that it is energetically
difficult to distinguish between a square and a triangular lattice.
This also means that it is quite difficult to differentiate between 
a lattice structure and a liquid phase, where the positions of
the vortices fluctuate but have suitable 
average distances; we do not introduce a different ansatz for 
the liquid case.  

We now parametrize 
the ansatz for the free energy of $n$ ($n\gg 1$)
type II vortices in an approximately 
triangular configuration measured on a square 
lattice ($L=L_1=L_2$) as
\be
{F(n)\over L_3e_3^2}=
{T\over e_3^2}
n\left[1+ XS_T(m_VL/\sqrt{n})\right],
\la{Fansatz}
\ee
where we allow eventually
$T,X,m_V$ to deviate from their mean field value. 
For $n=1$, we have instead one vortex
interacting with its replicas on an infinite periodic square lattice
of period $L=L_1=L_2$,
\be
\frac{T(L)}{e_3^2}={F(n=1)\over L_3e_3^2}=
{T\over e_3^2}
\left[1+ XS_S(m_VL) \right],
\la{Tansatz}
\ee
and this can be used for the 
finite size scaling studies of the tension $T$
and the interaction energy $X$.
For two vortices the period is $L/\sqrt2$,
\be
\frac{T_2(L)}{e_3^2}={F(n=2)\over 2L_3e_3^2}
=\frac{T}{e_3^2}\left[1+ XS_S(m_VL/\sqrt2) \right],
\la{T2ansatz}
\ee
which allows similarly to determine $T,X$.

{}From \eq\nr{Fansatz}
we can further, by differentiating with respect to $n$ in analogy 
with \eq\nr{dF}, derive that
\be
{H\over e_3^3}={dF(n)\over 2\pi L_3e_3^2}={T\over2\pi e_3^2}\left[
1+X\left(S_T(z)-\fr12 zS_T^\prime(z)\right)\right],
\la{hansatz}
\ee
evaluated at $z=m_VL/\sqrt{n}$.

\paragraph{The determination of $H_{c1}, H_{c2}$.}
When the volume and the 
number of vortices are large, the question is
what is their equilibrium configuration. Let us note first that
at small enough $H$, one is again in the Meissner phase
just as in the type I region. The 
critical value $H_{c1}$
for the transition to the vortex lattice phase can 
be obtained from \eq\nr{hansatz}: in the infinite volume limit, 
$S_T(z)\to 0$ and 
\be
\frac{H_{c1}}{e_3^3} = \frac{T}{2\pi e_3^2} \stackrel{\rmi{MF}}{=}
-\frac{y}{2x}{\cal E}(\sqrt{2x}).
\la{Hc1def}
\ee
The magnetic flux $B$ increases continuously 
from zero, unlike for type I where there is a discontinuity. 

Once $H>H_{c1}$, the vortices form
a triangular Abrikosov lattice structure
on the mean field level. Increasing
$H$ further on, the lattice structure finally disappears
when the energy squared corresponding to the lowest scalar 
Landau level becomes positive. This happens at
\be
H_{c2}^\rmi{MF}=-y\,e_3^3.
\la{hc2mf}
\ee
Note that the physics determining $H_{c1},H_{c2}$ is quite different
from that determining $H_c$ in the type I region, \eq\nr{hcmf}; 
at $x\to 1/2$ the expressions of course meet. 
 
\section{Fluctuations}
\la{phasediag}

The mean field approximation discussed above
misses some essential features of the
full path integral in \eq\nr{action}. For example, 
for a vanishing magnetic field and for small $x$
(type I region) the transition is of first order, 
while for large $x$ (type II region) the precise 
characteristics of the continuous transition have 
not yet been completely understood. These fluctuation effects
can be discussed perturbatively \cite{kleinert}, but must eventually
be studied with lattice simulations~\cite{u1big,u1}. 
There is no local order parameter to distinguish between 
the phases, but a number of non-local ones, e.g.,
the vector (photon) mass \cite{u1big} and 
the vortex tension \cite{tension}.

In general, the expansion parameters in the theory 
of \eq\nr{action} are, dimensionally, of the form 
(coupling constant)/(inverse correlation length $\equiv$ ``mass'').
Choosing the scalar coupling and the scalar mass, corresponds
to the standard Ginzburg criterion $(f_\rmi{symmetric}-
f_\rmi{broken})\times ({\rm scalar\, 
\,correlation\,\, length})^3$ $>T$. In
terms of the dimensionless variables $y,x$,
this means
\be
y<-16x^2.
\la{gi}
\ee
Thus, for mean field theory to be valid, one simply has to be 
sufficiently deep in the broken phase. However, this is 
not the only expansion parameter:
choosing instead, e.g., the vector coupling and the vector mass, 
$e_3^2/(\pi m_V)$, one gets a criterion which is more 
stringent at small $x$, 
\be
y < -10 x.
\ee
In any case, these are only qualitative criteria and the true size 
of fluctuation effects can only be established by numerical means.
In many analogous models, the fluctuations change the mean field
phase structure completely~\cite{phasestructure}.

For a non-vanishing field $B$, much new structure appears even on the
MF level, as discussed above. What happens in the full theory can,
in principle, be solved by the numerical methods presented here.
Many simulations have been carried out with different simplified models
(see, e.g., ~\cite{zt} and references 
therein), but it is not clear if the results should agree with
the full locally gauge invariant U(1)+Higgs theory, preserving
all relevant degrees of freedom.

Let us again consider the canonical ensemble \nr{canonical},
in which the magnetic field $H$ is fixed and $B$ is determined by
the system.
If the system is in the broken phase at $H=0$, and $H$
is increased, the system eventually changes to the symmetric phase,
which is characterized by a zero photon mass.
For type I, the transition is of first order, and no other phases
are believed to exist. 

For type II, the magnetic field penetrates
the system at strong enough fields, $H>H_{c1}=e_3T/2\pi$,
via vortices, and the photon becomes massless. 
On the mean field level, the result would be a vortex lattice. However, 
it is clear that this picture changes when fluctuations are included.
The interaction between the vortices becomes weaker when their distance
increases and therefore at small enough magnetic field, the fluctuations
will certainly destroy the lattice ordering.
On the other hand, when the field is large,
the vortices start to overlap and the cannot anymore be treated as
interacting line-like objects. 
There is experimental evidence \cite{nat} as well as 
theoretical understanding in other models~\cite{blatter,ffh,hsh} 
that the lattice existing at intermediate fields, then 
melts into a vortex liquid phase, which is 
thought to be smoothly
connected to the symmetric phase~\cite{ffh}. 

Therefore, it is an important question 
whether the lattice ordering predicted by the mean field theory
really exists at all for intermediate magnetic fields
in the U(1)+Higgs model. If there is no lattice phase, 
then the only true phase transition in the system is 
at $H_{c1}$, when the Meissner phase changes into a vortex liquid.

\begin{figure}[p] 

\vspace*{-1cm}

  \centerline{\vspace*{-4cm}
    \psfig{file=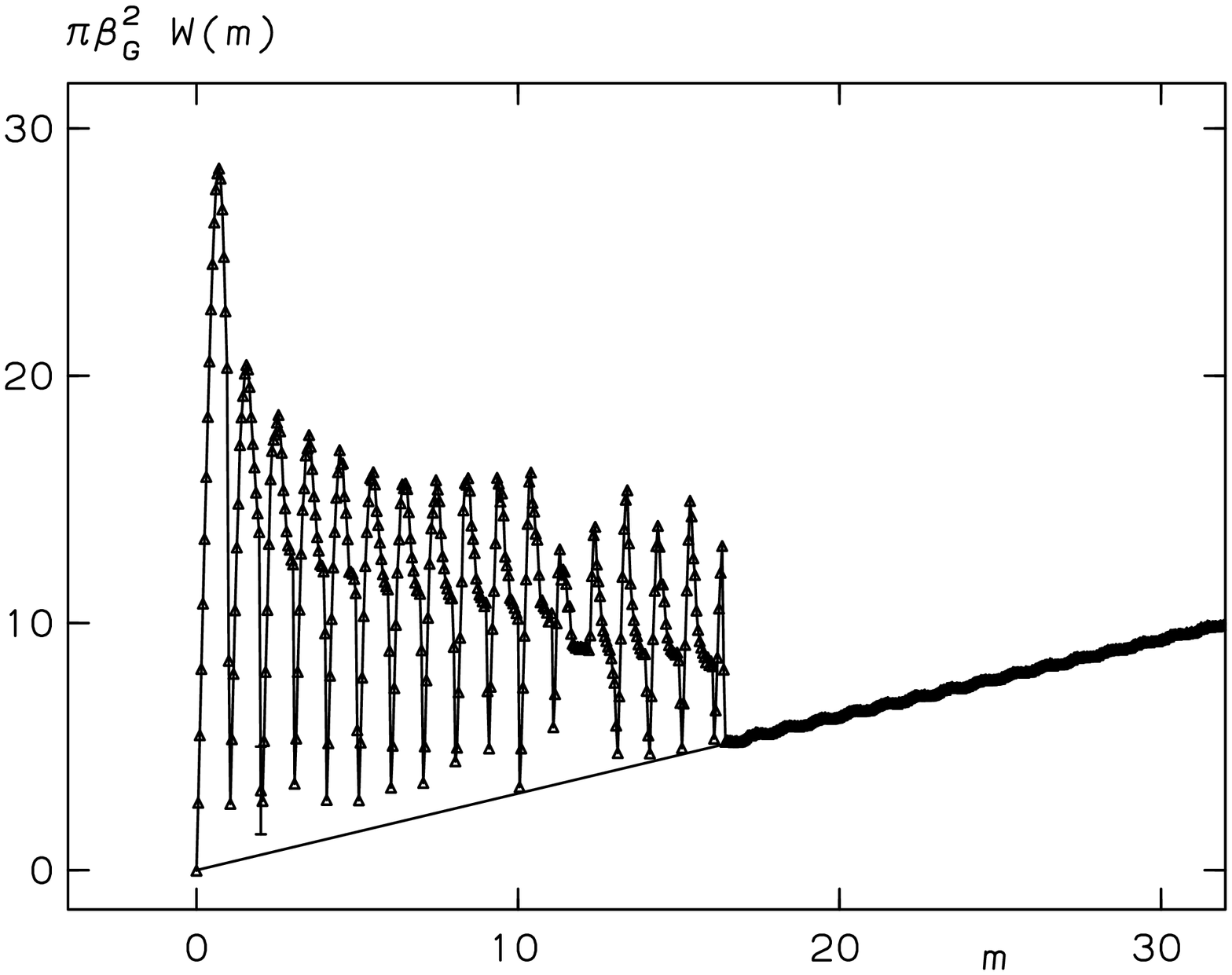,angle=270,width=10cm}
    }\hspace*{2cm}(a) 

\vspace*{2cm}

  \centerline{\vspace*{-4cm}
    \psfig{file=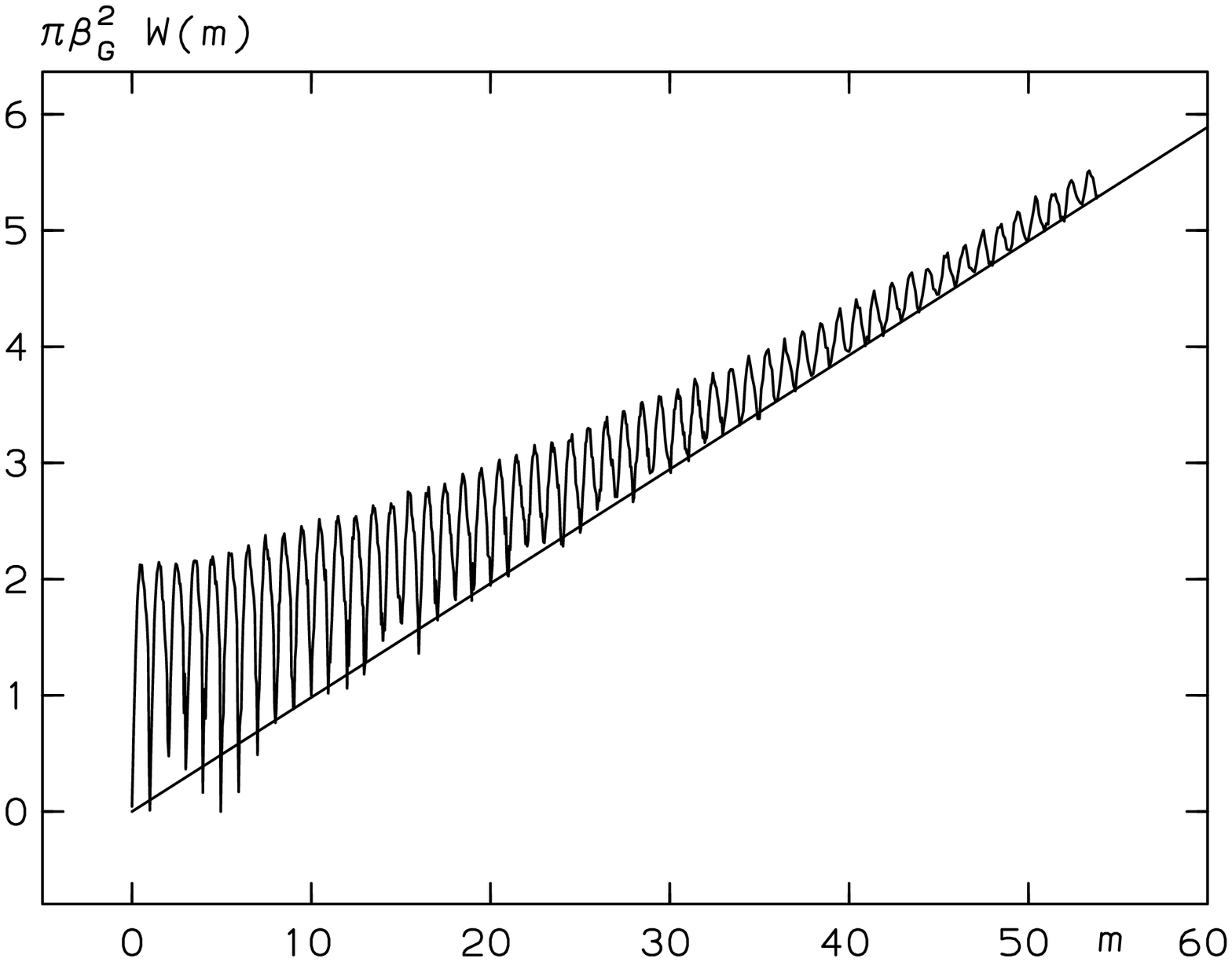,angle=270,width=10cm}
    }\hspace*{2cm}(b)

\vspace*{2cm}

  \centerline{\vspace*{-4cm}
    \psfig{file=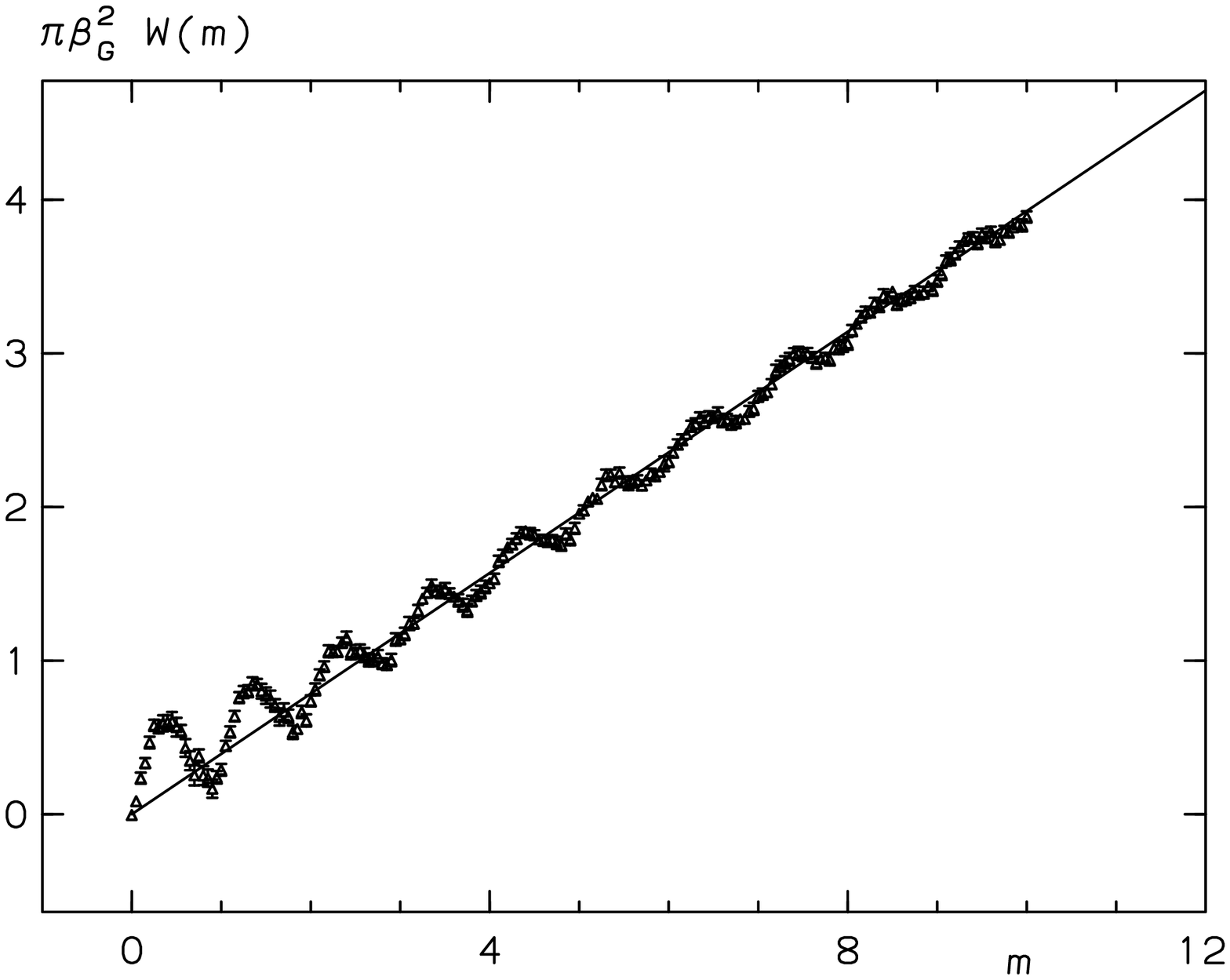,angle=270,width=10cm}
    }\hspace*{2cm}(c)

\vspace*{3cm}

\caption[t]{The function $\pi\beta_G^2W(m)$ (see \eq\nr{F(N)}) measured for 
(a) $x=0.0463$ (type I), $y=-2$, $\beta_G=4$, on a $18^2\times 16$ lattice;
(b) $x=2$ (type II), $y=-4$, $\beta_G=4$, on a $32^2\times 16$ lattice;
(c) $x=2$ (type II), $y=-1$, $\beta_G=4$, on a $16^3$ lattice.
The straight line is $W(m)=2m/(N_1N_2)$ (cf.\ \eq\nr{wmappro}).
}
\la{WtypeI}
\la{energy_type_ii}
\la{WtypeIIa}
\end{figure}

\section{Simulations in the type I region, $x<1/2$}
\la{sec:typeI}

The objective of the simulations in the type I region 
is to verify the qualitative picture discussed in Sec.~\ref{meanfield},
and to determine quantitatively the vortex tension $T$, the
critical field strength $H_c$, and the surface tension $\sigma$.
The simulations are carried out at one
fixed $x=0.0463$ and for $y=-1,-2,-4$. The lattice sizes are
$N_3=16,28$, $N_1=N_2=16,24,28,32$. The number of flux units
varies from 0 to about 33, corresponding to a range of
$B/e_3^3$ from 0 to about 10. The mean field values of the scalar
and vector correlation lengths are, at $\beta_G=4$, 
\ba
{1\over m_S^\rmi{MF}}&=&
{1\over\sqrt{-2y}}\,\beta_G a = 2a,\quad(y=-2),\nn
{1\over m_V^\rmi{MF}}&=&
\sqrt{{x\over -y}}\,\beta_G a \approx 0.6a,\quad(y=-2),
\nonumber
\ea
and indicate that the approach to the continuum limit 
must be checked explicitly (since the vector correlation
length is only of the order of the lattice spacing). 
{}From \cite{kp}, we know
that the tree-level estimates for the masses are
reasonably good everywhere in the broken phase. 

The numerical simulations, as presented here, were
performed with the use of hybrid Monte Carlo update schemes for the
gauge and the Higgs fields on parallel computer architectures. These
were supplemented by simulations on workstation clusters, where the gauge
fields were evolved using the more conventional $3$-hit Metropolis update.
A typical simulation for the calculation of the free energy change
$dF(n)$ consumed (1...several)$\times 10^5$ Monte Carlo sweeps
of either algorithm. Such values of the statistics allow for
relative errorbars at a few percent level in the measured
quantity. Errorbars have been calculated with jack-knife methods and
whenever necessary statistical error propagation was used.
For a more detailed account of the algorithms, error calculation 
and the calculation of integrals we refer to~\cite{tension}.

All the measurements are based on measuring the function $W(m)$ in
\eq\nr{F(N)}. Since all the analysis is based on its integrals,
i.e. free energies, we show only one example of $W(m)$ itself
in the type I region
in \fig\ref{WtypeI}(a);
all other cases are very similar. \fig\ref{WtypeI}(a)
exhibits in a striking way all the features associated with increasing
the number of flux units as described in Section \ref{meanfield}.
For smaller $m$, $W(m)$ contains a series of peaks between integer values
$m=n$. We know from \eq\nr{wmappro} 
that at each integer $W(m)=2m/N_1N_2$. This value
is reached within errorbars (not shown for clarity).

The peak in between the integer values
represents the addition of one more vortex line, and
the integral over the peak gives the additional free energy. The
clustering of flux lines as a cylinder (a bubble on the 
2-dimensional transverse
plane) is seen as a decrease of peaks (\eq\nr{Sbubble}), 
and the slab
as a relatively constant series of peaks (\eq\nr{Sslab}). At some
large $n$ the peaks suddenly disappear and the system goes over
to the symmetric phase. At this point it may
be illuminating to look also at the corresponding figures for the
type II region in \figs\ref{energy_type_ii}(b,c), which 
do not seem to show any abrupt transition.

\subsection{Vortex tension and interaction energy from $n$ = 1, 2}

We begin the study of the integrals of $W(m)$ by considering them
up to $n=1,2$. The vortex tension $T$ was already
studied in detail in \cite{tension}, and a comparison of $T$ and $T_2$
carried out here
is a measurement of vortex-vortex binding energy, which 
allows to differentiate between type I and type II regions. We choose 
$y=-1,-2$  and study finite size 
and finite lattice spacing effects in $T$ and $T_2$.
The values are obtained from \eqs\nr{T}, \nr{T2}.

\begin{figure}[tb]

\vspace*{-1cm}

  \centerline{ 
    \psfig{file=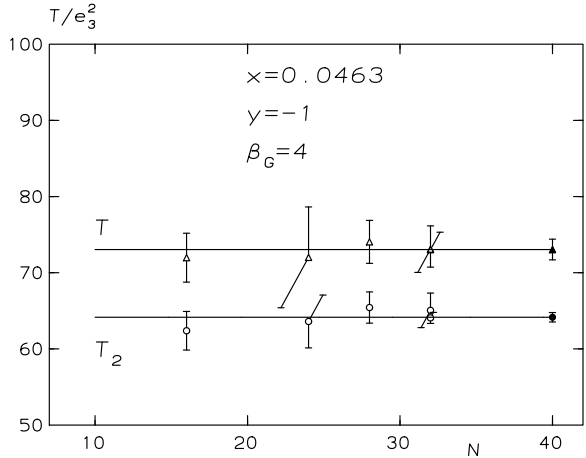,angle=270,width=10cm}
    } 

\vspace*{-0.5cm}

%
%

\caption[t]{ $T$ and $T_2$ as a function of $N_1=N_2$ at 
$\beta_G=4$, $x=0.0463$, $y=-1$. 
          }
  \label{t_of_l_463_at_1}
  \label{t_of_l_463_at_2}
\end{figure}

In Fig. \ref{t_of_l_463_at_1} we display $T$ and $T_2$ as a function
of $N_1=N_2$   
at $\beta_G=4$ and $y=-1$. The data
shows no finite size effects. Fitting to a constant value one obtains
\be
\begin{array}{rcll}
T/e_3^2&=&73.0(13) &(y=-1, \beta_G=4),\\
T_2/e_3^2&=&64.2(6) &(y=-1, \beta_G=4).
\end{array}
\ee
The fact that $T>T_2$ indicates an attractive force between the vortices.

\begin{figure}[tb]

\vspace*{-1cm}

  \centerline{
    \psfig{file=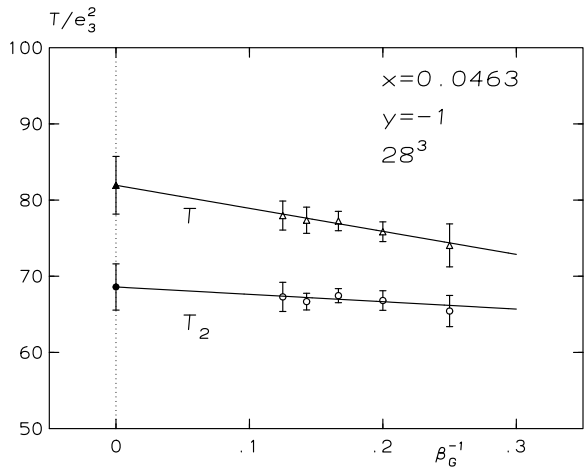,angle=270,width=10cm}
    }

\vspace*{-0.5cm}

  \caption{ $T$ and $T_2$ as a function of $\beta_G^{-1}$ on
a $28^3$ lattice and for $x=0.0463, y=-1$.
          }
  \label{t_of_b_463_at_1}
\end{figure}

To study the approach to continuum, we display
in \fig\ref{t_of_b_463_at_1} $T$ and $T_2$ as a function
of $\beta_G^{-1}=ae_3^2$ on a $28^3$ lattice at $y=-1$.
We can use a single lattice size since finite volume
effects were observed to be small. 
A dependence on $a$ is clearly seen and linear continuum
extrapolations (the straight lines in the figure) result in
\be
\begin{array}{rcll}
T/e_3^2&=&81.9(37) &(y=-1, \beta_G=\infty),\cr
T_2/e_3^2&=&68.6(30) &(y=-1, \beta_G=\infty).
\end{array}
\ee
According to \eq\nr{T} the mean field value at this $x$
is $T^\rmi{MF}=-42.75ye_3^2$ so that
$T/T^\rmi{MF}=1.91(8)$.
The magnitude of the scaling corrections is at the level of $10$\%
for $T$, somewhat smaller for $T_2$, if the continuum
results are compared with those at $\beta_G=4$. 

To study the $y$ dependence, we have measured
$T$ and $T_2$ 
also at $y=-2$. The data
again shows no apparent finite size effects. The result of a fit 
to a  constant value is 
\be
\begin{array}{rcll}
T/e_3^2&=&118.9(21),&(y=-2,\beta_G=4)\cr
T_2/e_3^2&=&106.2(8),&(y=-2,\beta_G=4)
\end{array}
\ee
corresponding to $T/T^\rmi{MF}=1.39(2)$.
Again, finite $\beta_G$ effects are expected to 
increase this number slightly. 

Summarising this subsection, one observes clearly 
a finite tension difference $T-T_2$ caused by the attractive
force between the vortices. There are also clear  
deviations from mean field which are largest close to the 
transition point, and actually quite sizable at $y=-1$.

\subsection{Results for large $n$}

\begin{figure}[p] 

\vspace*{-1cm}

  \centerline{\vspace*{-4cm}
    \psfig{file=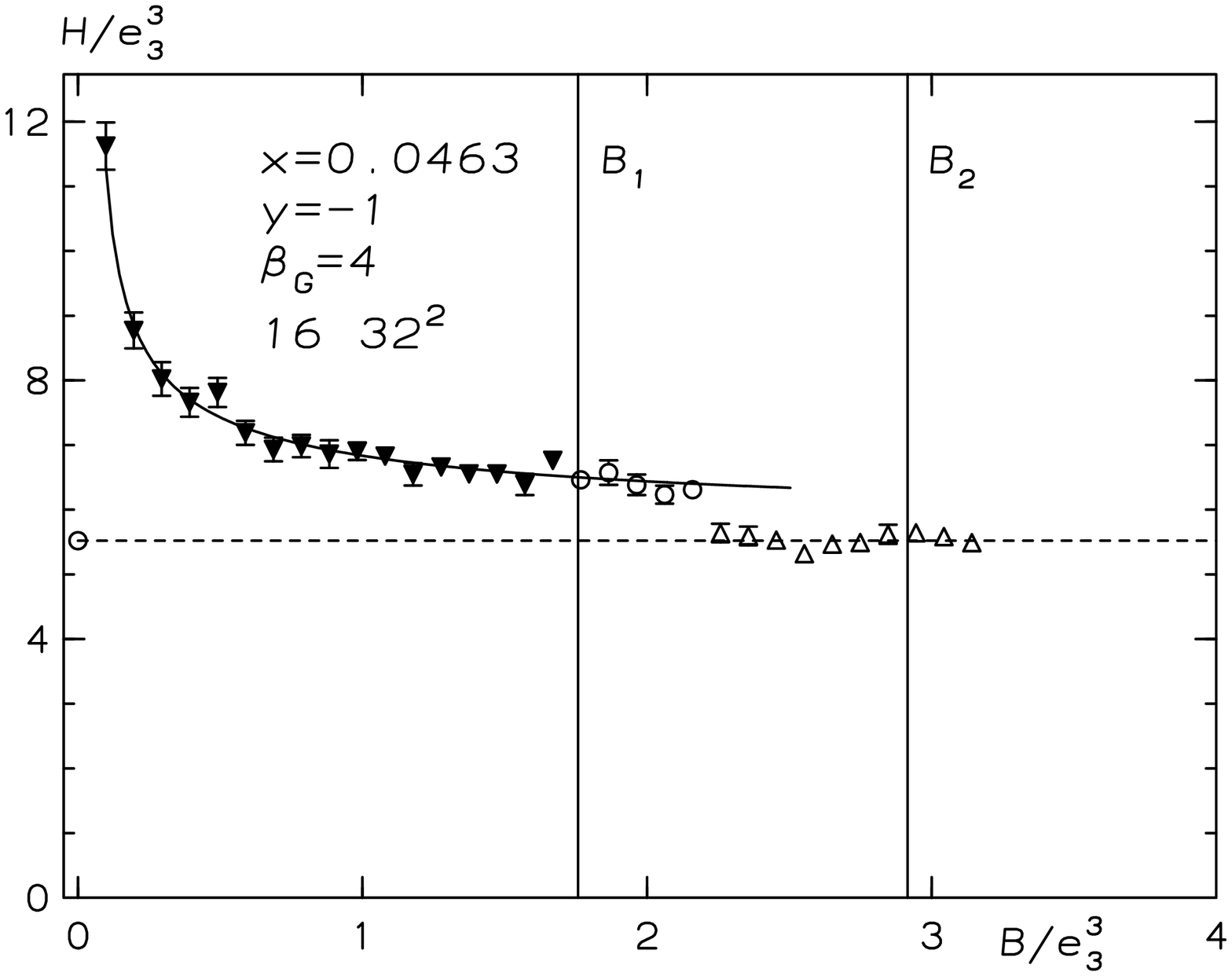,angle=270,width=10cm}
    }\hspace*{2cm}(a) 

\vspace*{2cm}

  \centerline{\vspace*{-4cm}
    \psfig{file=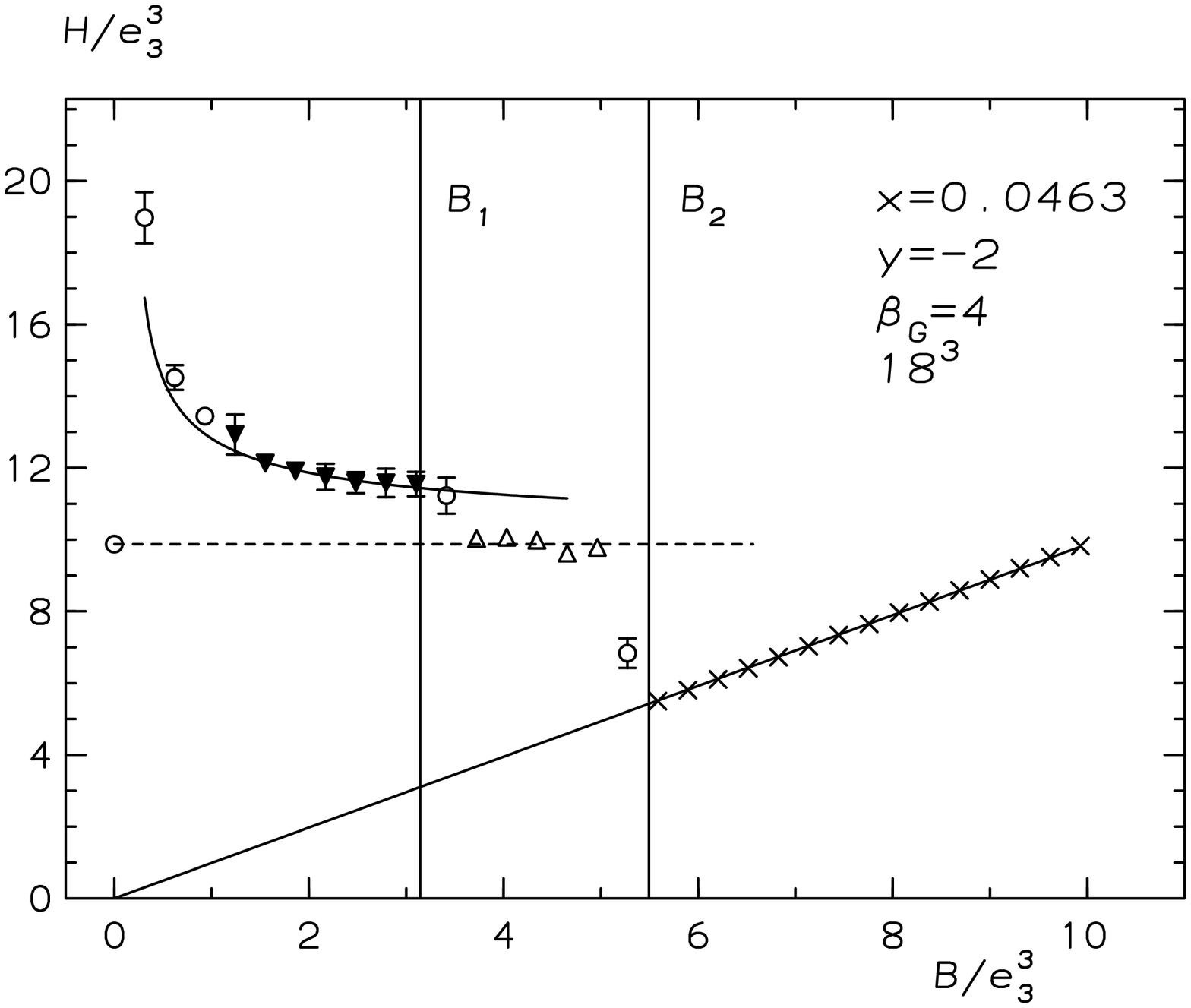,angle=270,width=10cm}
    }\hspace*{2cm}(b)

\vspace*{2cm}

  \centerline{\vspace*{-4cm}
    \psfig{file=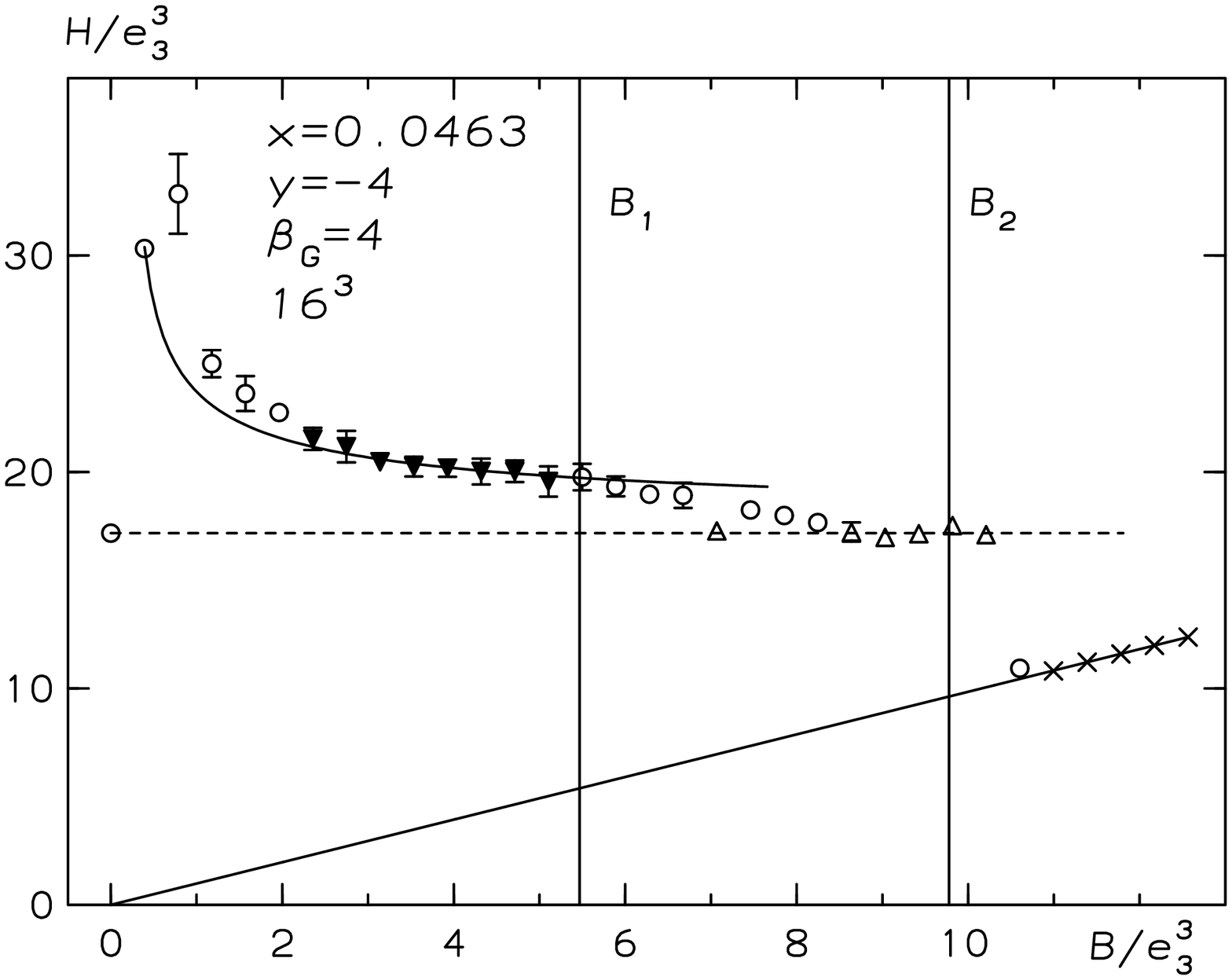,angle=270,width=10cm}
    }\hspace*{2cm}(c)

\vspace*{3cm}

\caption[t]{The magnetic field $H(B)$ (\eq\nr{dF}) for 
            $x=0.0463$ as a function of $B$ (\eq\nr{flux}): 
           (a) $y=-1$ on a $16 \times 32^2$ lattice;
           (b) $y=-2$ on a $18^3$ lattice;
           (c) $y=-4$ on a $16^3$ lattice.
           The flux number $n$ can be simply counted by starting 
           from the left. For vertical lines, see \eq\nr{b1b2}.
          }
  \label{t_1_1}
  \label{t_1_2}
  \label{t_1_4}
\end{figure}

Data for $H(B)$ (see \eq\nr{dF}) is presented 
in \figs\ref{t_1_1}(a--c) for
three sets of parameter values, 
$y=-1$ on a $16 \times 32^2$ lattice, $y=-2$ on a
$18^3$ lattice, and $y=-4$ on a $16^3$ lattice. We remind that this
data contains integration over the successive peaks of $W(m)$ 
such as those in \fig\ref{WtypeI}(a).

In all the figures, in agreement with Section~\ref{meanfield},
one can see a region with a rapid variation (cylinder region),
in which the data can be fitted to \eq\nr{dFbubble},
\be
H(B)=
H_c+\sigma\sqrt{\frac{\pi}{L_1L_2BH_c}},
\la{dFbubble2}
\ee
and a constant region (slab region), fitted to 
\eq\nr{dFslab}. The constant region gives the value of
$H_c$, the rapidly varying region the value of $\sigma$. After
these values are known, one can compute the two transition values:
$B_1$ in \eq\nr{bubbleslab} for the cylinder $\to$ slab transition
and $B_2$ in \eq\nr{critB} 
for the slab $\to$ symmetric transition:
\be
{B_1}={H_c \over\pi},\quad B_2=H_c-2e_3
\sqrt{{\sigma\beta_G\over N_1}}.  
\la{b1b2}
\ee
These are plotted in \figs\ref{t_1_1}(a--c) as vertical
lines. One observes that the flat section is
approximately located within the $B$-interval
$B_1<B<B_2$. For still larger lattices there would be a second cylinder
stage (\eq\nr{Sbubble3}), but now the condition \nr{skipcondition} is
satisfied and the transition takes place directly from a slab.

   The horizontal lines in all three figures correspond to fits to the flat
section. They determine the critical field $H_c$, which in the
mean field approximation is given by \eq\nr{hcmf}, 
$H_c^\rmi{MF}=-3.286ye_3^3$. 
We obtain
\be
\begin{array}{rccll}
H_c&=&5.52(3)e_3^3&\left(=1.680(10)H_c^\rmi{MF}\right) &(y=-1),\cr
&=&9.87(8)e_3^3&\left(=1.502(12)H_c^\rmi{MF}\right) &(y=-2),\cr
&=&17.17(9)e_3^3&\left(=1.306(07)H_c^\rmi{MF}\right) &(y=-4).
\end{array}
\la{typeI:Hc}
\ee

   The curves in the figures are fits to $H(B)$ for values
$B<B_1$ with the form \nr{dFbubble} and they determine $\sigma$.
At small $n$ one sees deviations from a spherical bubble shape and
correspondingly the fits use $y=-1$ data with 
$n \ge 1$, $y=-2$ data with $n \ge 4$ and $y=-4$ data with $n
\ge 6$. 
Compared with the value of $\sigma^\rmi{MF}=4.61(-y)^{3/2}e_3^4$ 
from \eq\nr{sigmainterface}, the fit results in
\be
\begin{array}{rccll}
\sigma&=&12.7(3)e_3^4&\left(=2.76(7)\sigma^\rmi{MF}\right) &(y=-1),\cr
&=&20.2(12)e_3^4&\left(=1.55(9)\sigma^\rmi{MF}\right) &(y=-2),\cr
&=&51.3(30)e_3^4&\left(=1.39(8)\sigma^\rmi{MF}\right) &(y=-4).
\end{array}
\ee

Finally, let us consider the thermodynamical 
discontinuities discussed in \eqs\nr{disc}. We note that 
at finite volumes, $\Delta B$ is directly given by the 
value ($\approx B_2$) where the system goes into the 
symmetric phase. However, there are large finite volume effects
in this quantity (cf.\ \eq\nr{critB}). In the infinite volume
limit, we expect rather that $\Delta B \approx H_c$, with the values
of $H_c$ as given in \eq\nr{typeI:Hc}.  Through \eqs\nr{latheat}, 
\nr{CC} and the curve $H_c(y)$ from \eq\nr{typeI:Hc}, 
this determines the latent heat.

Concluding this section, we observe that we have been able to 
determine $T-T_2$, verifying that we are indeed in the type I region, 
as well as all the main thermodynamical properties of the first 
order transition from the Meissner phase to the Coulomb phase. 
Qualitatively, the mean field
picture is valid, but on a quantitative level we observe a transition
which is significantly stronger than at the mean field level, 
particularly for small $|y|$. 

\section{Simulations in the type II region, $x>1/2$}
\la{sec:typeII}

  The simulations in the type II region have been
carried out at $x=2$,  $y=-0.4,\,-1,\,-4,\, -8$. 
The data for $y=-0.4,\,-1$ is mainly for the $n=1,2$ sectors
of the theory to study vortex-vortex interactions and their finite size
effects on periodic boxes. The data for $y=-4,\,-8$ is for
large winding numbers and
serves the purpose of determining the properties of the
theory at a finite field $H$.

As in the case of type I, all runs first compute $W(m)$ but the analysis
is based on its integrals. We thus present only two examples of 
$W(m)$ in \figs\ref{energy_type_ii}(b,c); the rest are similar and 
differ characteristically from those for type I in \fig\ref{WtypeI}(a).

The analysis then proceeds as follows:
\bi
\item To have a rough picture of the spatial structure of a
single vortex, we measure it in two ways: by explicitly measuring the
profile of the vortex and by measuring the vector correlation 
length, which is expected to set the scale of a type II vortex
at large distances.
\item To have an understanding of vortex-vortex interactions we
use finite size scaling of $T$ (\eq\nr{Tansatz}), the tension of a single 
vortex, and of $T_2$ (\eq\nr{T2ansatz}), the free energy per vortex of a
two-vortex system. We check that in the infinite volume limit $T=T_2$.
\item To confirm the understanding so obtained, we check whether 
the measured free energies of a many-vortex system,
$n=1,2,...,\sim50$, can be reproduced. We also attempt
to determine the phase of the system in the many-vortex region. 
\ei

\subsection{Vector field mass}
\la{vectormass}

To have information on the physical distance scales of the system,
we shall first for $n=0$ and in the broken phase determine the 
vector field $A_i$ mass $m_V$
(the ``photon mass''). We remind that
this quantity is a (non-local) order parameter for the transition:
it vanishes in the symmetric phase 
even after non-perturbative effects are taken into account~\cite{u1big}. 
Since we are
in the type II region, the scalar mass $m_S$ is always heavier 
than $m_V$ and is thus subdominant at large distances (for explicit 
measurements of $m_S$, see \cite{u1big}).

The vector mass is determined as follows. We define the plane
averaged correlator
\begin{equation}
{\cal O}_V(x_3;k_1)= \sum_{x_1,x_2}e^{ik_1x_1}\alpha_{12}(x_1,x_2,x_3),
\end{equation}
choosing the lowest non-zero value $k_1=2\pi/N_1$ for $k_1$. 
We then measure and fit the correlator as
\begin{equation}
\Gamma(ax_3)=
\langle{\cal O}(0;k_1){\cal O}^*(x_3;k_1)\rangle=
c(e^{-\omega x_3}+e^{-\omega (N_3-x_3)}).
\label{correlation_function_fit}
\end{equation}
The frequency mode $\omega$ is related to $m_V$ via 
the lattice dispersion relation
\begin{equation}
m_V^2a^2 = 
\Bigl[2\sinh(\frac{1}{2}\omega) \Bigr]^2-
\Bigl[2\sin(\frac{1}{2} k_1 )   \Bigr]^2.
\end{equation}

\begin{figure}[tb]

\vspace*{-1cm}

  \centerline{
    \psfig{file=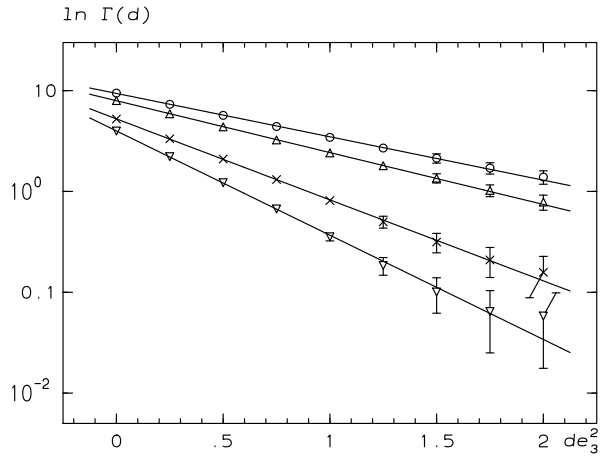,angle=270,width=10cm}
    }

\vspace*{-0.5cm}

  \caption{ Photon correlation functions on a $32^2\times 64$  
   lattice at $y=-8\,{\rm (bottom)},\,-4,\,-1,\,-0.4$ (top)
   as a function of $d=ax_3$.
   The curves correspond to the fit  in
\eq\nr{correlation_function_fit} with one massive scale. 
          }
  \label{correlation_function}
\end{figure}

We have measured $m_V$ for $y=-0.4,\,-1,\,-4,\,-8$ and for
lattices of geometry $N_1=N_2$ and  $N_3=2N_1$ with 
$N_1=16,\,24,\,32$. 
In \fig\ref{correlation_function},
examples of the correlation functions are displayed. 
A fit in accord with \eq\nr{correlation_function_fit}
gives a perfect description of the data, and higher states 
appear to be strongly suppressed. 

All values of $m_V$ determined here are collected in Table 
\ref{photon_mass}.
For comparison, the numerical 
value of the mean field mass $m_V^\rmi{MF}=e_3^2\sqrt{-y/x}$
is also given in the table.
The masses exhibit a mild and partly
non-monotonous dependence on $N_1$.
This means that we are not yet at large enough volumes, 
particularly for $y=-0.4,-1$. To have an estimate of 
the infinite volume values, we use a fit of the form
\begin{equation}
m_V(N_1)=m_V(N_1=\infty)+c\,e^{-m_V^\rmi{MF} N_1},
\end{equation}
assuming finite size effects to be those of a massive theory.
Other analogous fit forms give comparable results within errorbars.
In case of a non-monotonous $m_V(N_1)$,
only the two largest lattice sizes are included in this fit.
The boldfaced numbers in Table \ref{photon_mass} correspond to the
extrapolations. They will be used in the following sections of the paper.

\begin{table}[h]
\begin{center}
\begin{tabular}{||c|c|c|c||}\hline
 $y$ & $N_1$ & $N_3$ & $m_V/e_3^2$ \\ \hline
 -0.4 & 16 & 32 & 0.660(21) \\ 
 -0.4 & 24 & 48 & 0.756(17) \\ 
 -0.4 & 32 & 64 & 0.609(14) \\ 
 -0.4 & $\infty$ & 
$\infty$ & {\bf 0.508(54)} \\ 
 -0.4 & MF & MF & {\it 0.447} \\ \hline

 -1   & 16 & 32 & 0.915(16) \\ 
 -1   & 24 & 48 & 0.966(13) \\ 
 -1   & 32 & 64 & 0.893(11) \\ 
 -1   & $\infty$ &    
$\infty$ & {\bf 0.870(32)} \\ 
 -1   & MF & MF & {\it 0.707} \\ \hline
\end{tabular}
\hspace{5mm}
\begin{tabular}{||c|c|c|c||}\hline
 $y$ & $N_1$ & $N_3$ & $m_V/e_3^2$ \\ \hline
 -4 & 16 & 32 & 1.585(10) \\ 
 -4 & 24 & 48 & 1.676(10) \\ 
 -4 & 32 & 64 & 1.693(18) \\ 
 -4 & $\infty$ &    
$\infty$ & {\bf 1.684(16)} \\ 
 -4 & MF & MF & {\it 1.414} \\ \hline

 -8 & 16 & 32 & 2.170(8)  \\ 
 -8 & 24 & 48 & 2.246(10) \\ 
 -8 & 32 & 64 & 2.286(18) \\ 
 -8 & $\infty$ &    
$\infty$ & {\bf 2.256(16)}  \\ 
 -8 & MF & MF & {\it 2.000}  \\ \hline
\end{tabular}
\end{center}
\caption[photon_mas]{Results for the vector mass $m_V$. The values
in italics are the mean field (MF) results and the boldface
values are the infinite-volume extrapolations.}
\label{photon_mass}
\end{table}

\subsection{Profile of the vortex}

\begin{figure}[tb]

\vspace*{-1cm}

  \centerline{
    \psfig{file=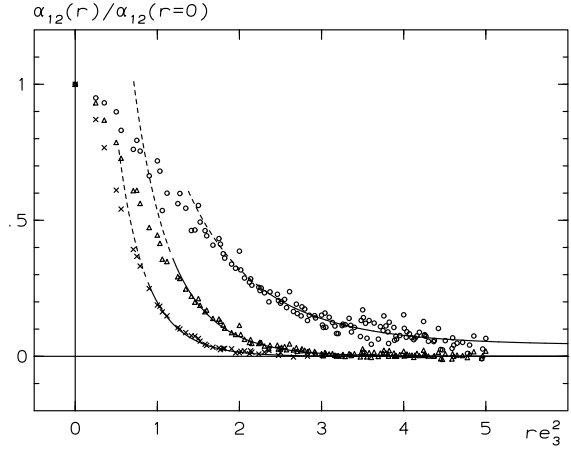,angle=270,width=10cm}
    }

\vspace*{-0.5cm}

  \caption{ The profile of the flux as a function of the distance $r$
from the center of the vortex. 
The curves ($y=-1$, top; $y=-4$ in the middle; $y=-8$,
bottom) correspond to fits according to 
\eq\nr{flux_fit_K0_kernel} for values of $z>2$. 
The parts of the fit curves corresponding to 
$z>2$ are drawn with solid lines.
             }
\la{vector_vortex_profile}
\end{figure}

The basic idea of our method of imposing a flux of $B$ on the system
is to put a defect along an arbitrary line and let the system dynamically
decide how it responds to this, where it places the vortices, etc. 
In the free energy measurement of a single vortex, its position never
enters. How can one then measure the profile of a vortex, i.e., 
the distribution of the flux density $\alpha_{12}$ on its planar 
cross section?

To do this one has to fix the position of the vortex. This can be done
by making the infinitely long defect in \eq\nr{equ:actchange} of finite
length in the $x_3$ direction. This effectively places 
a monopole and an antimonopole at the ends of the defect and localises
the response. Concretely, take a
$17\times17\times32$ lattice and shift the plaquette values by
$\alpha_{12}(x_1,x_2,x_3)\to\alpha_{12}(x_1,x_2,x_3)+2\pi$ for fixed
values $x_1=x_2=8$ but only for values of $x_3=8,9,...,23,24$. Then
the monopoles are located at 
$(8+\half,8+\half,7+\half)$ and $(8+\half,8+\half,24+\half)$, 
which gives a total
distance between the monopoles of $d_{mm}=17a=4.25/e_3^2$ in 
the $x_3$-direction. The Dirac string between the monopoles is again
compensated for by a vortex, and the
field profiles are measured at half-way distance 
between the monopoles, i.e.,\ at $x_3=16$. 

Let $r$ be the distance 
from the center of the defect $(8+\half,8+\half)$.
Assuming rotational invariance we measure the statistical average of the
quantity
\begin{equation}
{{\alpha_{12} (r)} \over {\alpha_{12}(r=0)}}.
\end{equation}
Results are displayed in \fig\ref{vector_vortex_profile} for 
$y=-8$. We have also repeated
the procedure for a $33\times33\times 64$ lattice and the data at 
$y=-4,\,-1$  in \fig\ref{vector_vortex_profile} stem from simulations
with that lattice size. 

In physical units, we expect the vortex size to be related to the
vector correlation length $1/m_V$.
We thus introduce  the dimensionless variable
\begin{equation}
z = m_V r,
\end{equation}
and we fit the large distance behavior of the vortex 
profile for values $z>2$
by the form
\begin{equation}
{{\alpha_{12} (z)} \over {\alpha_{12}(z=0)}}=c_1+c_2K_0(z).
\label{flux_fit_K0_kernel}
\end{equation}
We find that the asymptotic distance behavior at $z>2$ is 
perfectly described by the fit. Thus the asymptotic profile is 
governed by the vector mass alone. 
In the cases $y=-8,-4$ 
the constant $c_1$ can be chosen to be 
exactly zero, while at $y=-1$ it has a slightly
positive value, which we suspect to be a finite size effect.
Note that the mean field value for the scalar mass 
$m_S^{\rm MF}=\sqrt{-2y}e_3^2$ at $x=2$ exceeds the photon mass by a
factor of two, so that its contribution can be neglected.

Clearly, this analysis is only valid if 
the distance between the monopoles is larger than the 
width of the vortex, $d_{mm}>1/m_V$. The distance
should not be too large, either, or the fluctuations of the vortex
spread the fields and in the limit of infinite separation, the field
distribution becomes uniform. We estimate that this leads to a
Gaussian probability distribution of width $\sim\sqrt{d_{mm}/T}$
for the position of the center of the vortex, and therefore, if
$r>\sqrt{d_{mm}/T}$, this effect can be neglected. In our case
$z>2$ guarantees that this is satisfied.

\subsection{Vortex-vortex interactions for $n=1,2$}

As discussed in \se\ref{sub:typeII}, a determination 
of $T$, $T_2$ from \eqs\nr{T}, \nr{T2} is subject to much 
larger finite size effects in the type II case than in the 
type I case. (For an explicit comparison, 
see \figs\ref{t_of_l_463_at_1}, \ref{tt2100}.) Thus, we
have to employ the ansatz discussed in \se\ref{sub:typeII} 
to determine $T,T_2$. We denote the finite volume 
values by $T(L), T_2(L)$.

All of our data for $T(L)$ and $T_2(L)$ at $x=2,y=-1,\beta_G=4$ 
measured on cubic $N^3$ lattices ($L=Na$) is shown
in \fig\ref{tt2100} 
as a function of the lattice size $N$. 
The pattern is similar for other values of $y$.
We recall that the tensions
(\eqs\nr{T},\nr{T2}) are obtained by measuring $W(m)$ (\eq\nr{F(N)})
and integrating (\eq\nr{dF}) over one or the two first peaks of
functions such as those in~\fig\ref{energy_type_ii}.

\begin{figure}[tb]

\vspace*{-1cm}

  \centerline{ 
    \psfig{file=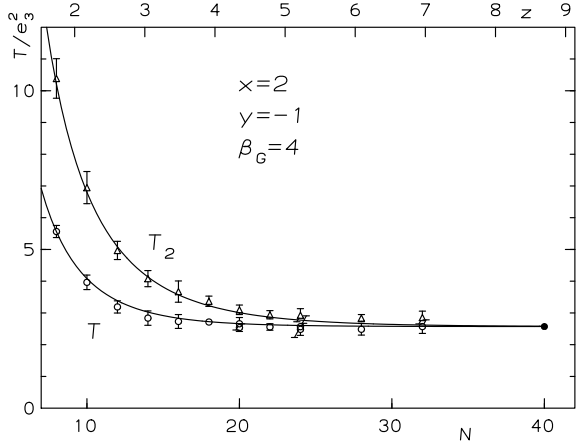,angle=270,width=10cm}
    } 

\vspace*{-0.5cm}

%
%

\caption[t]{$T_2(L)$ (triangles) and $T(L)$ (circles)
at $x=2,\,\beta_G=4$ and $y=-1$, 
measured on an $N^3$ lattice as a function of $N$. 
             }
\la{tt2100}
\la{tt0400}
\end{figure}

The data in \fig\ref{tt2100} 
is then fitted
with two free parameters, namely $T$ and $X$ according to the finite
size scaling laws in
\eqs\nr{Tansatz} and \nr{T2ansatz}, using $m_V$ as determined 
in \se\ref{vectormass}. The numerical value of the
distance control variable $z$ is, for $N_1=N_2=N$,
\ba
z=m_VL
&=&0.564N \quad(y=-8),\nn
&=&0.421N \quad(y=-4),\nn
&=&0.218N \quad(y=-1),\nn
&=&0.127N \quad(y=-0.4).
\la{zvalue}
\ea
We include data with $z\gsim1$ in the
lattice sums in \eqs\nr{lattsums}. 
Typical finite size scaling studies
in other lattice models use values of $z>2$, but we have checked that
our fit results are stable under further omission of small $z$ data.
The fits for $T$
(such as the lower curve in \fig\ref{tt2100})
have $\chi^2_\rmi{dof}=0.48$
at $y=-1$ and $\chi^2_\rmi{dof}=0.69$ at $y=-0.4$.
The parameters are
\be
\begin{array}{rcrcl}
T&=17.87(35)e_3^2\left(=1.077(21)T^\rmi{MF}\right),&
X&=... \hspace*{1.3cm} & (y=-8),\cr
&=9.331(91)e_3^2\left(=1.125(11)T^\rmi{MF}\right),&
&=...  \hspace*{1.3cm}& (y=-4),\cr
&=2.571(39)e_3^2\left(=1.240(19)T^\rmi{MF}\right),&
&=1.02(6)[10] & (y=-1),\cr
&=1.089(30)e_3^2\left(=1.313(36)T^\rmi{MF}\right),& &=0.80(3)[20] 
&(y=-0.4),
\end{array}
\label{fit_100_a}
\ee
where the mean field result is
$T^\rmi{MF}=-2.0735ye_3^2$, 
and the numbers in the square brackets correspond to 
changing the (infinite volume) 
values of $m_V$ within the errorbars
given in Table~\ref{photon_mass}.
For $y=-8,-4$, the data is not good enough for determining $X$, 
while for all values of $T$, the errors from the 
uncertainty in $m_V$ are within the statistical errorbars. 
Both $T(L)$ and $T_2(L)$ give comparable values 
for the infinite volume extrapolation (see \fig\ref{tt2100}).
It is quite impressive that such large
finite size corrections are described so well by the fit. This is
a clear indication of the fact that the interaction is indeed described
by $K_0$. 

Consider then the approach to continuum, $a=1/(\beta_Ge_3^2)\to0$, 
of $T$ and $T_2$. We choose $y=-1$.
For a fixed $N^3=28^3$ lattice we determine $T$ and $T_2$
at $\beta_G=1,2,4,6$ and
$8$. Table \ref{100_continuum} contains in its second
and fourth column data for the measured quantities $T(L)$ and $T_2(L)$,
$L=28a$. This data is then extrapolated to infinite volume using
the previous measurements of $X$  in  
\eq\nr{fit_100_a} and $m_V$ in Table~\ref{photon_mass},
and assuming the validity of \nr{Tansatz} and
\nr{T2ansatz}. The extrapolations are given in columns $3$ and $5$ of
Table \ref{100_continuum}. Note that we have assumed that $X$, $m_V$
do not depend significantly on $\beta_G$, and we use the same 
values for each $\beta_G$; this clearly introduces some extra 
$\beta_G$-dependence, but it turns out that the outcome of this rough
procedure is numerically quite satisfactory. 
The data somewhat scatters; at large values of $\beta_G$ 
autocorrelations within the Monte Carlo process are large.

\begin{figure}[t]

\vspace*{-1cm}

  \centerline{
    \psfig{file=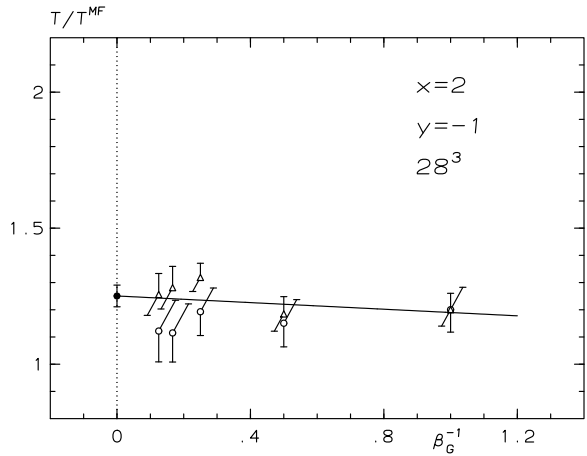,angle=270,width=10cm}
    }

\vspace*{-0.5cm}

  \caption{The continuum extrapolation of $T/T^\rmi{MF}$ 
    and $T_2/T^\rmi{MF}$ 
    as a function of $1/\beta_G$ at $x=2$ and $y=-1$. 
    Here $T^\rmi{MF}=2.0735 e_3^2$.
    Circles and
    triangles correspond to infinite volume extrapolations of
    $T/T^\rmi{MF}$ and $T_2/T^\rmi{MF}$. 
    The solid circle and the fitted line correspond to the
    continuum extrapolation.
    }
\la{100_continuum_plot}
\end{figure}

\begin{table}[ht]
\begin{center}
\begin{tabular}{||c|c|c|c|c||}\hline
 $\beta_G^{-1}$ & $T(L)/e_3^2$ & $T(L=\infty)/e_3^2$ & 
$T_2(L)/e_3^2$ & $T_2(L=\infty)/e_3^2$ \\ \hline
 1.0000 & 2.488(170) & 2.488(170) & 2.489(125) &  2.489(124) \\ 
  .5000 & 2.385(179) & 2.385(179) & 2.457(131) &  2.457(131) \\ 
  .2500 & 2.485(182) & 2.473(181) & 2.838(112) &  2.735(107) \\ 
  .1666 & 2.428(232) & 2.312(221) & 3.248(198) &  2.656(162) \\ 
  .1250 & 2.738(276) & 2.326(237) & 4.187(256) &  2.605(159) \\ \hline
\end{tabular}
\end{center}
\caption[100_continuum]{The infinite volume ($L=\infty$)
extrapolations of $T/e_3^2$ and $T_2/e_3^2$
as a function of $\beta_G^{-1}$ at $x=2$ and $y=-1$; $L=28a$ (see text).
For $\beta_G^{-1}=1.0,0.5$, the dependence on $L$ is within the errorbars.}
\label{100_continuum}
\end{table}
 
We display
$T(L=\infty)$ and $T_2(L=\infty)$ as
a function of $ae_3^2=1/\beta_G$ in \fig\ref{100_continuum_plot}.
The straight line fit (the curve in the figure) corresponds to the form 
\begin{equation}
T(\beta_G)=T+{c\over{\beta_G}}   
\end{equation}
for $T$ and a similar one for $T_2$. The fit results in
\begin{equation}
T= 2.594(83)e_3^2\left(=1.251(40)T^\rmi{MF}\right) \quad (y=-1),
\end{equation}
having $\chi^2_\rmi{dof}=0.83$.
This value is completely compatible with the one determined in
\eq\nr{fit_100_a}
at $y=-1$ and thus we conclude that finite-$a$ corrections
are small and apparently
under control for the values of $\beta_G$ considered.
\fig\ref{100_continuum_plot} can be compared with that 
for the type I case, \fig\ref{t_of_b_463_at_1}.

To summarize this subsection, we have verified that at $x=2$, 
$T-T_2 < 0 $ at finite volumes but $\to 0$ for $L\to\infty$.
Hence we are in the type II region. We then showed that 
the infinite volume extrapolations $T(L=\infty)$ can in turn be 
extrapolated to a finite physical value in the continuum limit. 

\subsection{The vortex system at large $n$}

Consider then the case of large $n$. As in the type I region, 
we plot $H(B)$ from \eq\nr{dF}.
Our main data is shown in
\fig\ref{fit_triangle_32}, which contains the integrals over
the peaks in $W(m)$, shown in \fig\ref{energy_type_ii}.

A first-order phase transition means that $B$ behaves 
discontinuously as a function of $H$.
In the microcanonical picture, it would
correspond to a region in which $H$ is constant,
or more precisely, would exhibit similar finite-size effects 
as those discussed in Sec.~\ref{typeImf}. 
The only region where this might be the case, is at small $B$
(cf. \fig\ref{fit_triangle_32}(b)). 
If that turns out to be true, 
there is a first-order phase transition
from $B=0$ or a small non-zero value, to 
some finite larger value of $B$. In principle the transition 
could be one {\em from a vortex liquid phase} at such small $B$  
that the distance between the vortices is large enough
that the interaction cannot preserve the lattice structure, 
to the lattice phase; or, more likely in a finite volume, {\em from 
a vortex lattice phase} enforced by the periodic boundary conditions, 
to a vortex liquid phase where the vortices are so close to each other
that the average fluctuations become larger than the average distance. 
  
\begin{figure}[p] 

\vspace*{-1cm}

  \centerline{\vspace*{-4cm}
    \psfig{file=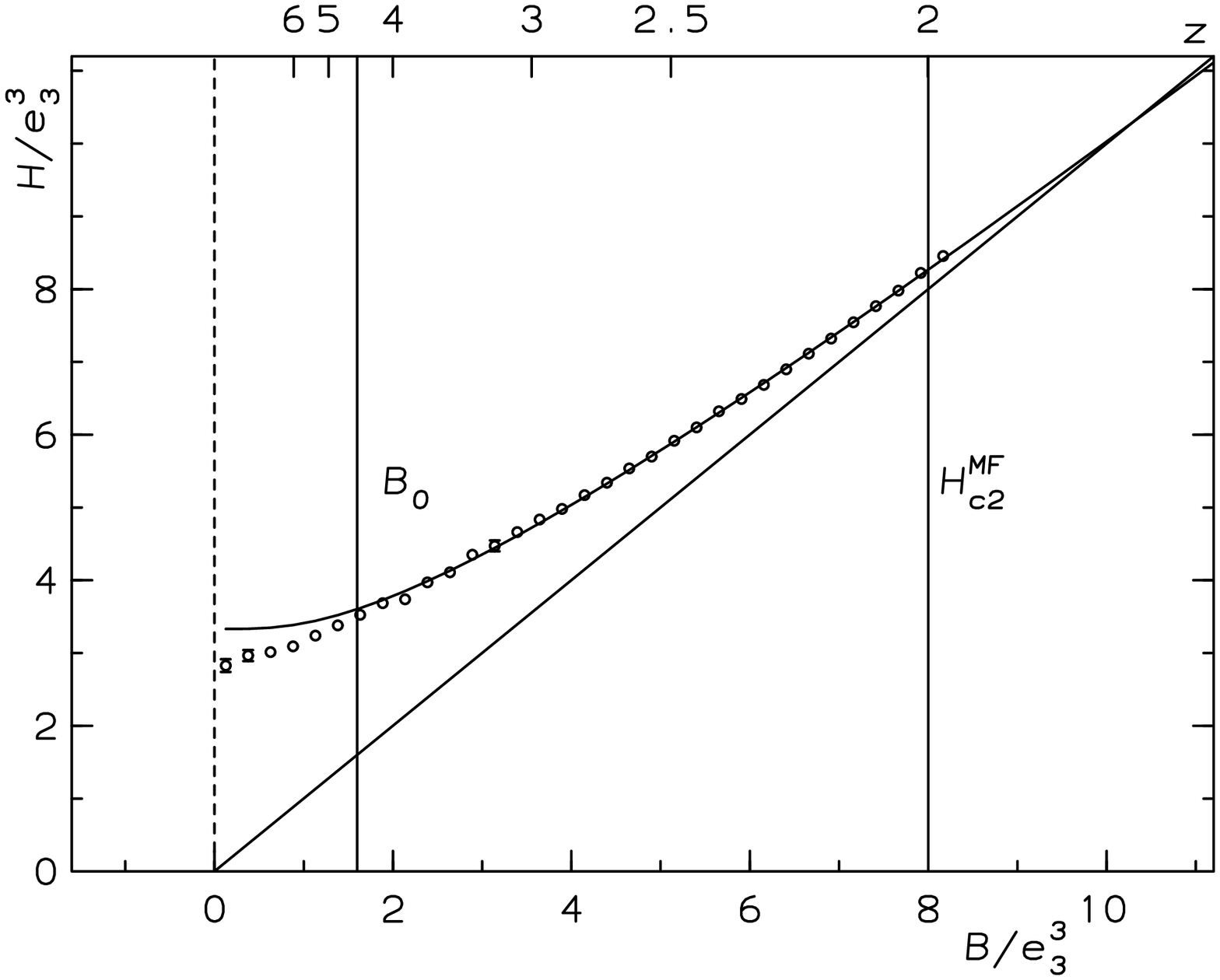,angle=270,width=10cm}
    }\hspace*{2cm}(a)

\vspace*{2cm}

  \centerline{\vspace*{-4cm}
    \psfig{file=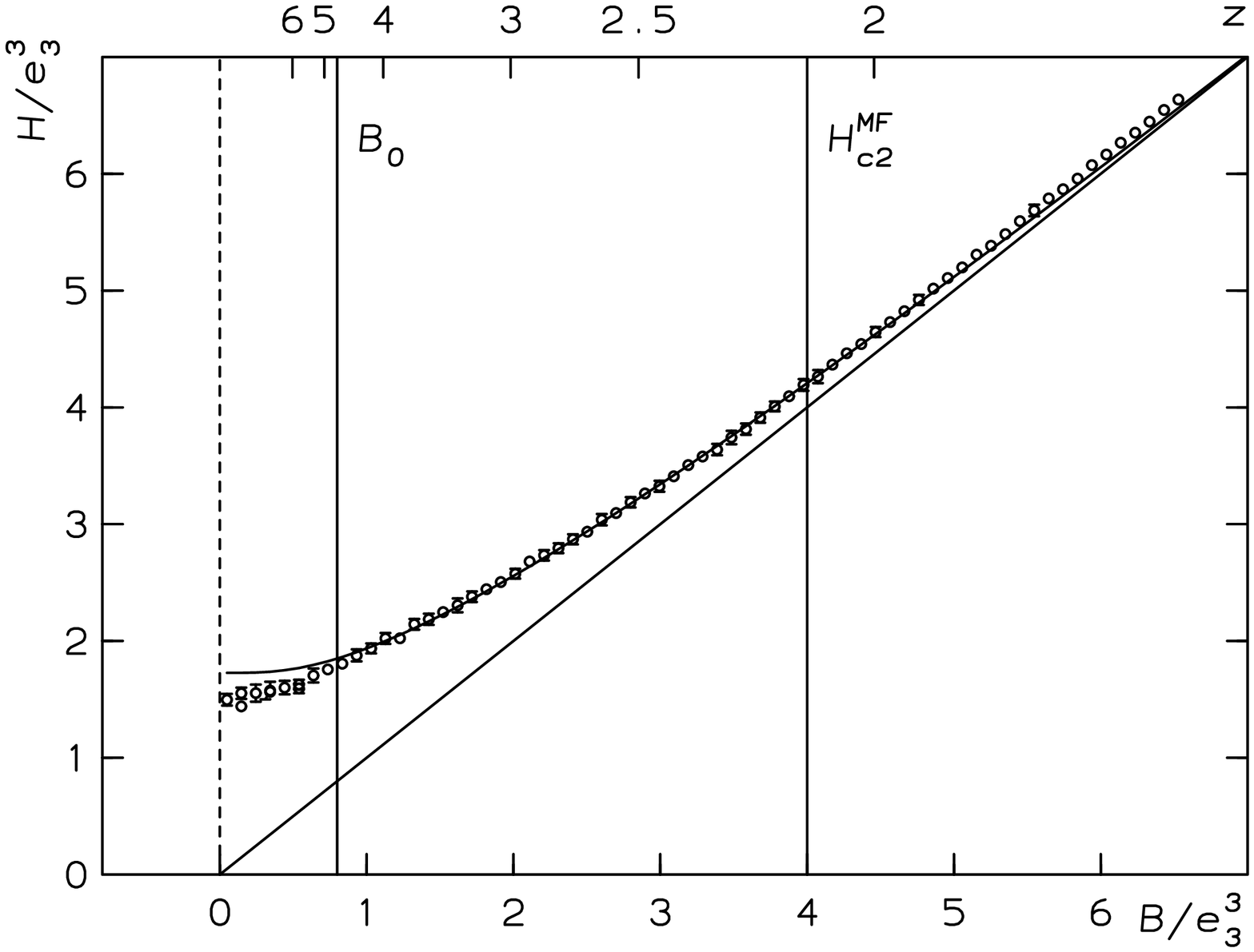,angle=270,width=10cm}
    }\hspace*{2cm}(b) 

\vspace*{2cm}

  \centerline{\vspace*{-4cm}
    \psfig{file=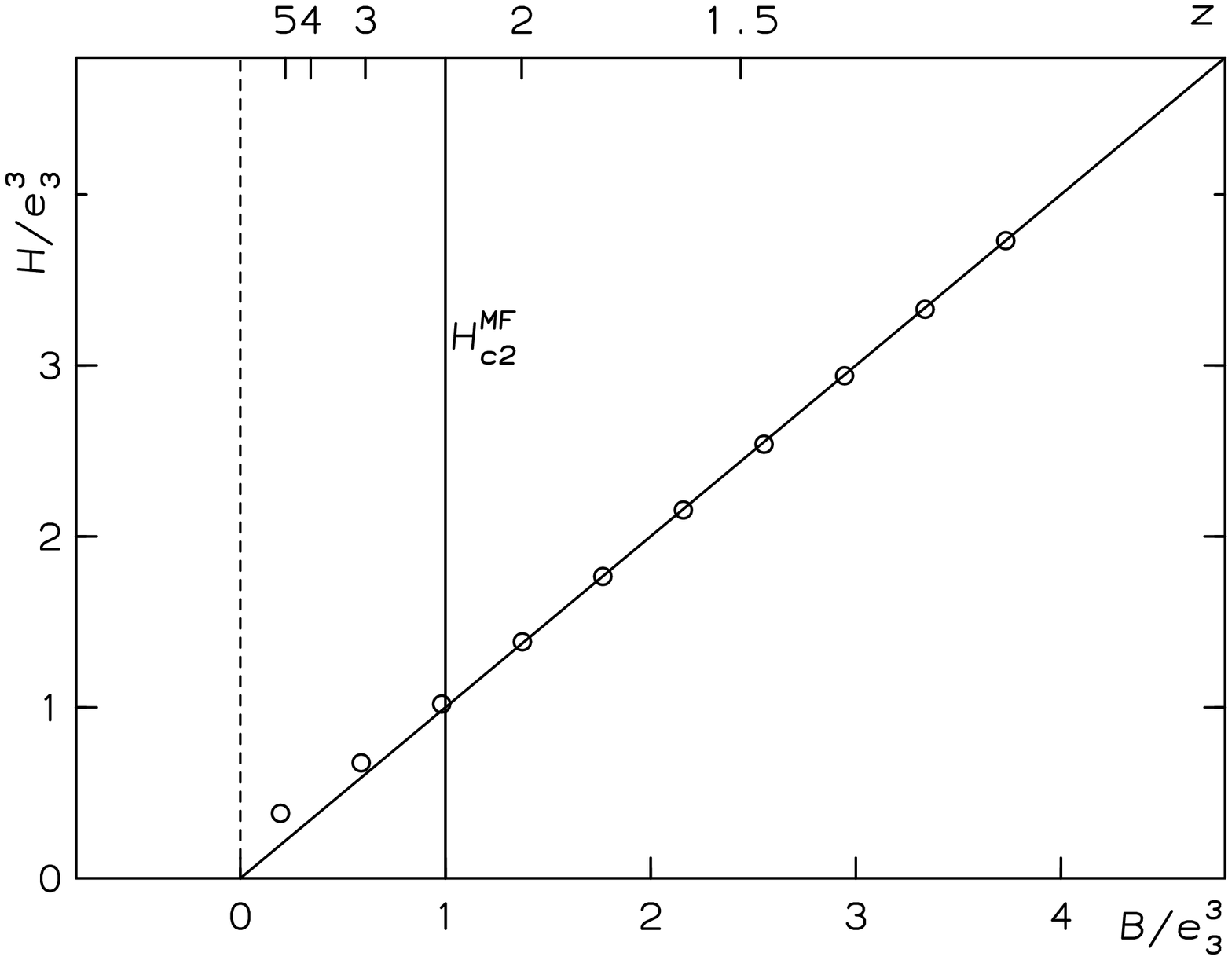,angle=270,width=10cm}
    }\hspace*{2cm}(c)

\vspace*{3cm}

\caption[t]{The field strength $H(B)$ as a function of 
            $B$ at $x=2.0$, $\beta_G=4$ and  
           (a) $y=-8$ on a $20^2\times16$ lattice;  
           (b) $y=-4$ on a $32^2\times16$ lattice; 
           (c) $y=-1$ on a $16^3$  
            lattice. The vertical lines in (a),(b) indicate the
            fit intervals as denoted by $B_0$ and $H^{\rm MF}_{c2}$
            for the triangular lattice fit in \eq\nr{hansatz}. As
            discussed in the text, the good agreement of the triangular
            fit with the data is not enough to guarantee that there
            really is a lattice structure in the system.  
    }
\la{fit_triangle_32}
\la{fit_triangle_20}
\la{fit_triangle_16}
\end{figure}

In order to check whether a lattice or a liquid phase 
is more likely, the data for $H(B)$ is fitted to the ansatz 
in \eq\nr{hansatz} with both a
triangular (T) and a square (S) lattice. We remind
that now the ansatz is for $H=dF/(VdB)$ and that the distance control
variable $z$ quantitatively is ($L_1=L_2=L=Na$)
\ba
z={m_VL\over\sqrt{n}}
=m_V\sqrt\frac{2\pi}{e_3B}
&=&{5.655\over\sqrt{B/e_3^3}}
\left(=\frac{11.28}{\sqrt{n}}\right) \quad (y=-8),\\
&=&{4.221\over\sqrt{B/e_3^3}}
\left(=\frac{13.46}{\sqrt{n}}\right)
\quad (y=-4),\nn
&=&{2.181\over\sqrt{B/e_3^3}}
\left(=\frac{3.49}{\sqrt{n}}\right)
\quad\,\,\, (y=-1),\nn
&=&{1.273\over\sqrt{B/e_3^3}}
\hspace*{2.2cm}  (y=-0.4), \nonumber
\ea
where the numbers in the parentheses correspond to 
the lattice sizes used in \fig\ref{fit_triangle_32}.
The fit is done for $y=-4,-8$ with 
a fixed interval $B_0<B<H_{c2}^\rmi{MF}$, 
where $B_0(y=-4)=0.8e_3^3$, $B_0(y=-8)=1.6e_3^3$ and
$H_{c2}^\rmi{MF}=-y\,e_3^3$. We consider $T$ and $X$ 
as free parameters in the fit. The results for the parameters
are given in Table \ref{fit_results} and the fits are shown 
in \figs\ref{fit_triangle_32}(a,b). 

\begin{table}[h]
\begin{center}
\begin{tabular}{||c|c|c|c|c|c||}\hline
$y$ & $N_1=N_2$ & $\Lambda$ & $X$ & $T/T^{MF}$ & $\chi^2_\rmi{dof}$ \\ \hline

-4 &    16 &   T &   0.91(06)  &    1.26(2)  &    0.32\\ 
-4 &    24 &   T &   0.90(05)  &    1.27(2)  &    0.09\\ 
-4 &    32 &   T &   0.88(04)  &    1.31(1)  &    0.33\\ 
-4 &    48 &   T &   0.96(14)  &    1.31(4)  &    0.18\\ \hline 

-4 &    16 &   S &   0.91(06)  &    1.25(2)  &    0.31\\ 
-4 &    24 &   S &   0.91(05)  &    1.26(2)  &    0.08\\ 
-4 &    32 &   S &   0.89(04)  &    1.30(1)  &    0.30\\ 
-4 &    48 &   S &   0.96(15)  &    1.31(4)  &    0.18\\ \hline 

-8 &    16 &   T &   0.81(05)  &    1.22(1)  &    0.42\\ 
-8 &    20 &   T &   0.78(05)  &    1.26(2)  &    0.79\\ 
-8 &    24 &   T &   0.93(10)  &    1.19(3)  &    0.24\\ \hline 

-8 &    16 &   S &   0.82(05)  &    1.22(1)  &    0.38\\ 
-8 &    20 &   S &   0.79(04)  &    1.25(2)  &    0.74\\ 
-8 &    24 &   S &   0.93(10)  &    1.18(3)  &    0.24\\ \hline 

\end{tabular}
\end{center}
\caption[fit_results]{Fit results for $X$ and $T$
from the fits in \figs\ref{fit_triangle_32}(a,b), 
with $0.8e_3^3<B<H_{c2}^\rmi{MF}$ ($y=-4$) 
and $1.6e_3^3<B<H_{c2}^\rmi{MF}$ ($y=-8$),
assuming the existence of a triangular (T) 
or a square (S) vortex lattice. The mean field 
value in the dilute limit is $X=0.76$.}
\label{fit_results}
\end{table}

To summarize the results for the parameters at different $y$-values:
for the vortex-vortex interaction parameter $X(y)$ we have
$X(-8)=0.80(5),\,\,X(-4)=0.90(5)$.
There is only mild $y$-dependence
and the results are consistent with
those at $y=-1,-.4$ in \eq\nr{fit_100_a}. 
The mean field value (applicable in the dilute limit)
is $X=0.76$, independent of $y$. The tension values 
$T(y)$ of the fit at large $B$ are
$T(-4)=10.7(3)e_3^2$, $T(-8)=19.9(5)e_3^2$.
These values differ from the
infinite volume tension as determined 
in the dilute limit in \eq\nr{fit_100_a} by
about 10 percent, which may well be a finite volume effect. 
Finally, the $S$ and $T$ lattice fits coincide
within errorbars. 

The fact that the $S$ and $T$ lattice fits work equally 
well and give consistent results, means that we are not 
sensitive to the average spatial distribution of the vortices. 
Thus, a vortex liquid phase cannot be excluded, either. 

To get more information on the phase, 
let us then inspect the region of large $B$.  
At $y=-4$ and $y=-8$, the peaks in $W(m)$ persist at all values of $n$
studied (see Fig.~\ref{energy_type_ii}), 
and the field strength $H(B)$ does not reach the
mean field Coulomb phase value $H(B)=B$, corresponding to 
vanishing magnetization $M(B)=B-H(B)=0$.
On the mean field level the transition to $M(B)=0$ 
takes place according to \eq\nr{hc2mf} at
\be
B=H_{c2}^\rmi{MF}= -y\,e_3^3,
\la{hc2}
\ee
which is shown in \figs\ref{fit_triangle_32}(a-c)
as a solid line and is well within the studied range. 
However, we do not observe any structure around these values. 

For the data set at $y=-4$, we have performed
a finite size scaling study of $M(B)$
at a large value of 
$B/e_3^3=6.2831$. Considering hypercubic $N=N_1=N_2=N_3$ boxes 
for values $N=8,12,16,20,24$, we display  
data for $M(B)$ as a function
of $N$ in Fig.~\ref{m_at_large_b}. 
The data differs from the mean field prediction $M(B)=0$
and suggests an infinite volume extrapolation to a value
$M(B/e_3^3=6.2831)\approx -0.12$. The fitted curve in the figure
corresponds to a finite size scaling ansatz
\be
M=M(B)+{A \over {N^\gamma}},
\la{fss_ansatz}
\ee
with $\gamma=1$. Such power law finite size corrections are expected
in a massless phase. The exponent $\gamma$ could in principle be 
non-trivial, but we are currently not in the position to determine
its value from the fit. In any case, it is clear that the deviation
of the measured $M(B)$ from the mean field value $M=0$ is not
a finite size effect.

\begin{figure}[t]

\vspace*{-1cm}

  \centerline{
    \psfig{file=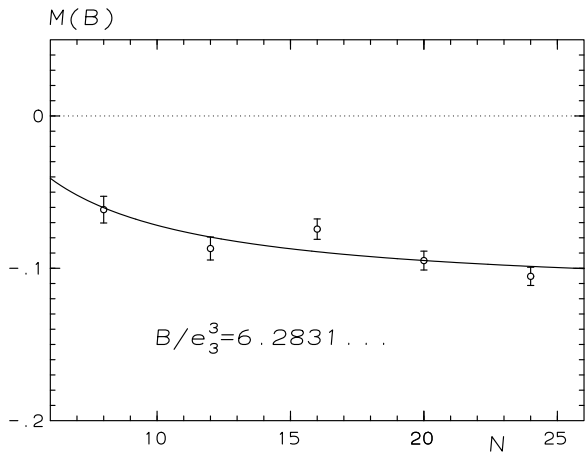,angle=270,width=10cm}
    }

\vspace*{-0.5cm}

  \caption{Finite size scaling analysis of $M(B)=B-H(B)$ at
   $x=2, y=-4$ and $B/e_3^3=6.2831$
   on $N^3$ boxes (compare with \fig\ref{fit_triangle_20}(b)). 
   The fitted curve corresponds to the ansatz
   in \eq\nr{fss_ansatz}. It extrapolates to a non-zero
   value of $M(B)$ here.
    }
\la{m_at_large_b}
\end{figure}

When $B$ is further increased, $H(B)$ should finally approach the
value $B$, but achieving high enough $n$ is costly in computer time.
The required value of $n$ can be reduced by, say, decreasing $-y$.
As an example, the $y=-1$ case  is shown
in \figs\ref{WtypeIIa}(c) and \ref{fit_triangle_16}(c). Here
$H(B)$ reaches $B$ within errorbars 
already at $n\approx 3$ but the data is not good for any
quantitative analysis. However, it illustrates an interesting point:
contrary to what one might conclude on the basis of the plots
for $W(m)$ in \figs\ref{WtypeIIa}(a,b), 
on finite lattices there are no cusps at integer values
of $m$; in addition, $W(m)$ can fall below the line $2m/(N_1N_2)$. 
In fact, $M=B-H(B)\approx 0$ is obtained when positive and negative 
contributions cancel out.

In experiments with high-temperature superconductors \cite{nat} a weakly
first-order phase transition has been found at a large magnetic field $H$, 
and it has been interpreted as a melting transition of the vortex lattice. 
Like the transition we could imagine to observe at small $B$,
it would  also show up in our data as a constant part in 
Fig.~\ref{fit_triangle_32}. Clearly, there is no sign of such 
a transition in our data, so if one exists in the U(1)+Higgs 
model at the value $x=2$ we have studied, the transition must be 
extremely weak. A much more natural conclusion seems 
to be that at least 
for the parameter values studied, the 3d U(1)+Higgs model
does not have a vortex lattice phase as a thermodynamical state, 
but that the structure obtained at the mean field level is 
removed by thermal fluctuations. Thus we would be 
in the liquid phase all the time, and 
it just smoothly changes into the symmetric Coulomb phase.
This is in agreement with the behaviour of $m_{V}$ as a
non-local gauge invariant order
parameter: there is only one ``Meissner" phase at $H<H_{c1}=
T/(2\pi)$ with $m_{V}\neq 0$, while
for $H>H_{c1}$, $m_{V}$ vanishes.

\section{Conclusions and outlook}
\la{concl}

\begin{figure}[t]


\centerline{\epsfxsize=8cm\hspace*{-1cm}\epsfbox{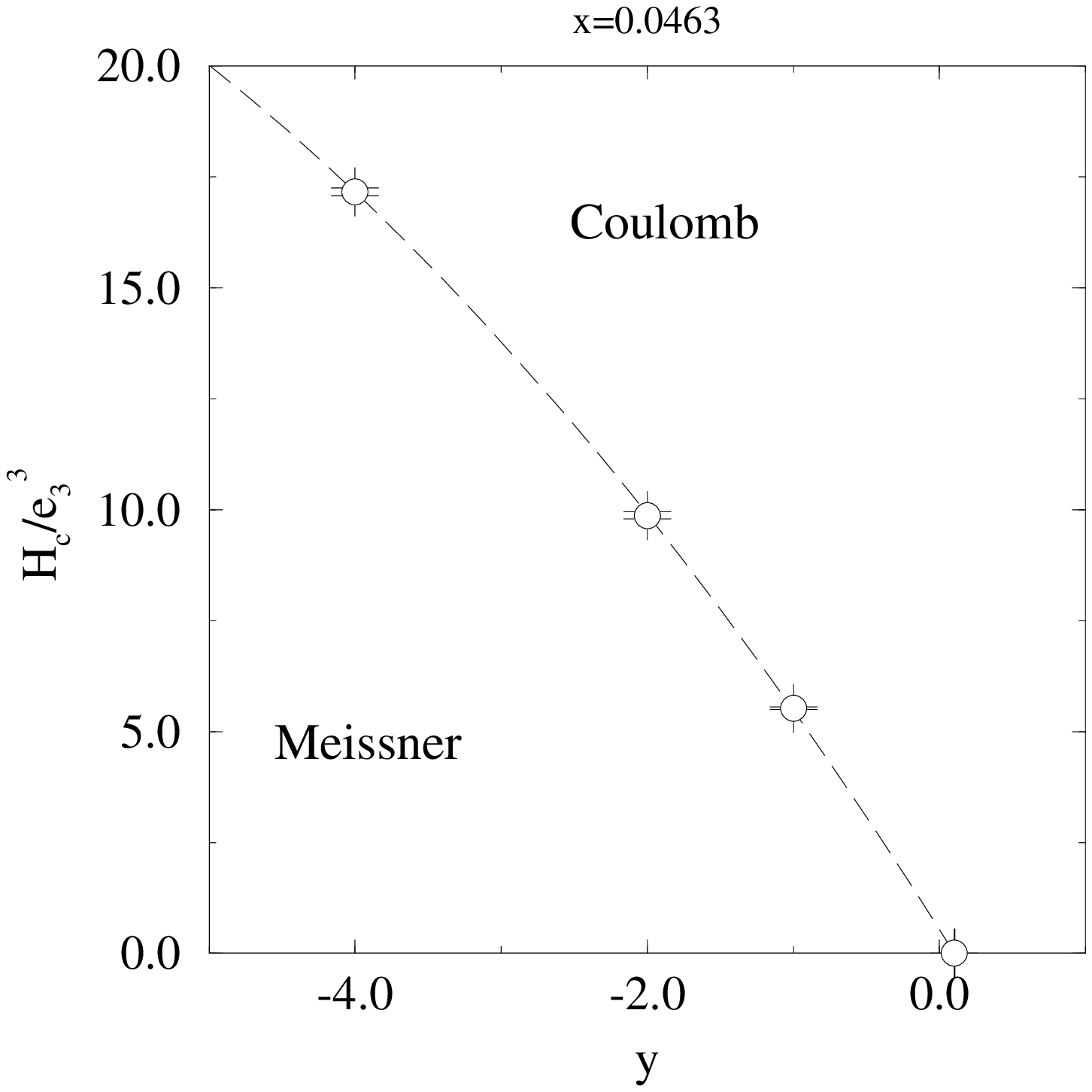}%
\epsfxsize=8cm\hspace*{-1cm}\epsfbox{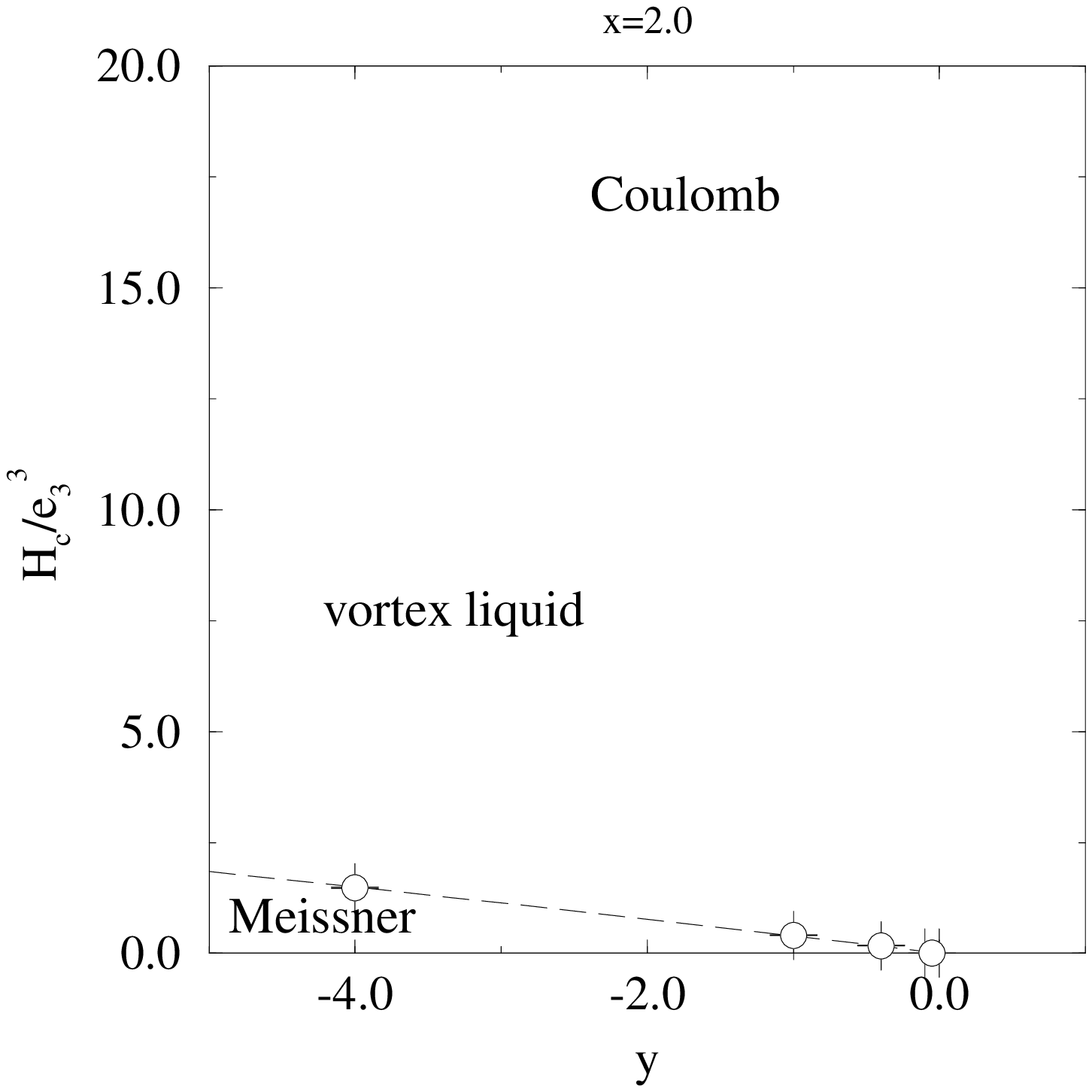}}

\vspace*{-4cm}

\caption[a]{The phase diagrams for $x=0.0463$ (left) and
$x=2.0$ (right). The critical values of $H_c$ in the type I region
are from \eq\nr{typeI:Hc}, for $H_{c1}$ in the type II region
from \eq\nr{fit_100_a} (cf.\ \eq\nr{Hc1def}), and for
$y$ at the point $H=0$ from~\cite{u1big}.} 
\la{fig:pd}
\end{figure}

In this paper, we have shown how the interactions of
vortices can be studied non-perturbatively with lattice
simulations, starting directly from field theory. We 
have shown that the interaction energy between two
vortices is negative in the type I region and positive
in the type II region, as expected from mean field (MF)
theory. Moreover, increasing the magnetic field
further on, we have shown that vortex interactions
can lead to phases with different
types of macroscopic structures.

The present simulations were carried out in a region
in which one approaches the limits of validity of 
MF-theory. In superconductivity $-y\gg1$ (see \eq\nr{ab})
and MF-theory works very well. Our simulations span the range
$y=-8,...,-0.4$ and we, indeed, find increasing deviations 
from MF. 

The regions of the phase diagram displaying 
Meissner, vortex liquid, vortex lattice, 
and normal Coulomb behaviour, have to date not been accurately mapped 
out in the full Ginzburg-Landau theory. We have demonstrated that the 
methods introduced here can answer these questions. Our 
present results and conclusions  
for the phase diagram in the $(y,H)$-plane at $x=0.0463,x=2$
are shown in \fig\ref{fig:pd}. At $x=0.0463$, we find a first
order phase transition in qualitative accordance with the  
MF estimate. However, the transition is much stronger than 
at the MF level, particularly for small $H_c$ (in fact, 
at the MF level the transition is of the 2nd order for 
$H_c\to 0$, while the actual transition continues to be of the 
first order there). The vector field mass is non-zero only
in the Meissner phase.

In the type II case $x=2$ and at $y=-4$, we observe only a single 
transition from the Meissner phase to a vortex liquid phase, 
which then smoothly turns into the Coulomb phase. This is 
in strong contrast to the MF prediction: we do not find 
any indications for a phase transition associated with 
the melting of a vortex lattice phase at large $B\approx H_{c2}^\rmi{MF}$,
so it seems natural to conclude that we are in the liquid 
phase all the time, and the fluctuations are strong enough 
to remove the lattice structure predicted at the MF level. On the other 
hand, we found very preliminary indications
(\fig\ref{fit_triangle_32}(b)) that the transition to the 
vortex liquid phase at small $B$ might be of the first order.
A possible interpretation for this behaviour is a finite
volume effect associated with the 'melting' of an
unphysical lattice structure induced by the periodic 
boundary conditions, for a small number of flux quanta
(in other words, a transition from a boundary-effect-dominated
region to a bulk-dominated region). 

Whether or not the theory in the thermodynamical limit
allows for a physical triangular vortex lattice state
in a strict sense at large values of $-y$, 
currently is an open question. It remains an 
algorithmic and computational challenge to extend the present simulations to 
the case of a truly macroscopic number of vortices.

One could imagine that 
techniques analogous to those introduced
here can be developed for defects other than vortices, as well.
For instance, in a theory with monopoles, one could fix
the total magnetic flux emerging from the whole lattice volume.
This would allow for a gauge-invariant 
non-perturbative study of the free energy of
one monopole or of an ensemble of several monopoles.

Finally, let us note that the present techniques can
be easily extended to the 3d SU(2)$\times$U(1)+Higgs 
theory, relevant for the cosmological electroweak
phase transition in the Standard Model and many 
extensions thereof. In that case, an external 
magnetic field can represent physical (cosmological) reality, 
and can in principle lead to the emergence of phases analogous
to the vortex lattice of type II superconductors~\cite{ao}.
Such phases might be relevant for baryogenesis and cosmology,
and would thus be of considerable interest. We have observed 
that even in the U(1)+Higgs system where the lattice prediction 
is much more robust to begin with, fluctuations tend to remove the 
structure and result in a vortex liquid phase. This makes it 
quite understandable that in the SU(2)$\times$U(1) case no lattice
structure (nor, in fact, a clear liquid-like behaviour) could be
observed in the region of the parameter space studied
so far~\cite{bext}, but one rather finds a symmetric phase.


\section*{Acknowledgements}

We acknowledge useful discussions with 
T. Ala-Nissil\"a, M.A. Moore, A. Sudb{\o},  
Z. Te{\v s}anovi{\' c} and M. Tsypin. 
Most of the simulations were carried out with a Cray T3E at the Center 
for Scientific Computing, Finland, and with a number of workstations
at the Helsinki Institute of Physics. Some simulations were also 
performed at NIC in J\"ulich. 
The total amount of computing power used corresponds to
about 5$\times 10^4$ hours of a single Cray node's capacity, 
i.e., $2\times 10^{16}$ floating-point operations. 
This work was partly supported by the 
TMR network {\em Finite Temperature Phase
Transitions in Particle Physics}, EU contract no.\ FMRX-CT97-0122.
A.R. was partly supported by the University of Helsinki.


\end{document}